\documentstyle[emulateapj,psfig,flushrt,tighten]{article}

\def\kms{\,{\rm km\,s^{-1}}}

\def\msun{\,{\rm M_\odot}}

\def\etal{{et al.\ }}

\newcommand\beq{\begin{equation}}
\newcommand\eeq{\end{equation}}
\newcommand{\ba}{\begin{eqnarray}}
\newcommand{\ea}{\end{eqnarray}}  
\def\spose#1{\hbox to 0pt{#1\hss}}
\def\lta{\mathrel{\spose{\lower 3pt\hbox{$\mathchar"218$}}
     \raise 2.0pt\hbox{$\mathchar"13C$}}}
\def\gta{\mathrel{\spose{\lower 3pt\hbox{$\mathchar"218$}}
     \raise 2.0pt\hbox{$\mathchar"13E$}}}

\righthead{Massive black hole binaries}
\submitted{ApJ, accepted}
\makeatletter
\newenvironment{tablehere}
  {\def\@captype{table}}
  {}
\newenvironment{figurehere}
  {\def\@captype{figure}}
  {}
\makeatother

\begin{document}

\title{Interaction of massive black hole binaries with their stellar environment: III. Scattering 
of bound stars}

\author{Alberto Sesana\altaffilmark{1,2}, Francesco Haardt\altaffilmark{1^*}, \&
Piero Madau\altaffilmark{3}}

\altaffiltext{1}{Dipartimento di Scienze, Universit\'a dell'Insubria, via 
Valleggio  11, 22100 Como, Italy. $^*$Affiliated to INFN, Italy.}
\altaffiltext{2}{Center for Gravitational Wave Physics, Penn State University, University Park, PA 16802, USA.}
\altaffiltext{3}{Department of Astronomy \& Astrophysics, University of 
California, 1156 High Street, Santa Cruz, CA 95064.}

\begin{abstract}
We develop a formalism for studying the dynamics of massive black hole binaries 
embedded in gravitationally-bound stellar cusps, and study the binary orbital decay 
by three-body interactions, the impact of stellar slingshots on the density profile of 
the inner cusp, and the properties of the ejected hypervelocity stars (HVSs). We find that
the scattering of bound stars shrinks the binary orbit and increases its eccentricity more effectively 
than that of unbound ambient stars. Binaries with initial eccentricities 
$e\gtrsim 0.3$ and/or unequal-mass companions ($M_2/M_1 \lesssim 0.1$) can decay by three-body 
interactions to the gravitational wave emission regime in less than a Hubble time.
The stellar cusp is significantly eroded, and cores as shallow as $\rho\propto r^{-0.7}$ 
may develop from a pre-existing singular isothermal density profile. A population of HVSs is 
ejected in the host galaxy halo, with a total mass $\sim M_2$. We scale our results to  
the scattering of stars bound to Sgr A$^*$, the massive black hole in the Galactic Center,
by an inspiraling companion of intermediate mass. Depending on binary mass ratio, eccentricity, 
and initial slope of the stellar cusp, a core of radius $\sim 0.1$ pc typically forms 
in 1-10 Myr. On this timescale about 500-2500 HVSs are expelled 
with speeds sufficiently large to escape the gravitational potential of the Milky Way.
\end{abstract}


\section{INTRODUCTION}

In the standard paradigm of cosmic structure formation, it is expected  that many wide massive 
black hole binaries  (MBHBs) will form following the mergers  of two massive galaxies 
(e.g. Begelman, Blandford, \& Rees 1980; Volonteri, Haardt, \& Madau 2003; Mayer \etal 2007).  
The binary will subsequently shrink due to stellar or gas dynamical processes and may 
ultimately coalesce by emitting a burst of gravitational waves. It was 
first proposed by Ebisuzaki, Makino, \& Okumura (1991) that the heating
of the surrounding stars by a decaying SMBH pair would create a low-density
core out of a preexisting cuspy (e.g. $\rho\propto r^{-2}$) stellar profile.
In a purely stellar background a `hard' binary shrinks by capturing the stars that pass
close to the holes and ejecting them at much higher velocities, a super-elastic
scattering process (`gravitational slingshot') that depletes the nuclear region. 
Observationally, there is evidence in early--type galaxies for a systematically 
different distribution of surface brightness profiles, with faint ellipticals showing
steep power-law profiles (cusps), while bright ellipticals have much
shallower stellar cores (e.g. Faber \etal 1997; Ravindranath \etal 2001). 
Dwarf ellipticals seem to elude this paradigm showing somewhat flat profiles, similarly to bright ones 
(Graham \& Guzman 2003). Late--type spirals tend to show steep central cusps, 
as in the case of the Milky Way or Andromeda. 
Detailed N-body simulations have confirmed the
cusp-disruption effect of a hardening MBHB (Makino \& Ebisuzaki 1996;
Quinlan \& Hernquist 1997; Milosavljevic \& Merritt 2001), while semi-analytic
modelling in the framework of hierarchical structure formation theories has
shown that the cumulative damage done to a stellar cusps by decaying 
black hole pairs may explain the observed correlation between the
`mass deficit' (the mass needed to bring a flat inner density profile to 
a $r^{-2}$ cusp) and the mass of the nuclear black hole (Volonteri, Madau, \& Haardt 2003).

This is the third paper of a series aimed at the detailed study of the interaction 
between MBHBs and their stellar environment. In Sesana, Haardt, \& Madau (2006,2007a, 
hereafter Paper I and Paper II), we analyzed the 
three-body scatterings between a MBHB and background stars {\it unbound} to the binary. 
The assumption of a fixed background breaks down once the binary has ejected most of 
the stars on intersecting orbits, and the extraction of energy and angular momentum 
from the binary can continue only if new stars can diffuse into low-angular momentum 
orbits (refilling the binary's phase-space ``loss cone''), or via gas processes.
In galaxies with inner cores or shallow cusps, only a small fraction of the loss cone 
is confined within the sphere of influence of the binary, and the  
approximation of a background of unbound stars is reasonable. 
A similar argument holds also in the case of a galaxy with a steep cusp 
hosting a nearly equal-mass binary ($M_1\sim M_2$). The radius of influence
of such a pair, $r_{\rm inf}=G(M_1+M_2)/(2\sigma^2)$ where $\sigma$ is the stellar
velocity dispersion, is 
of the order of the binary hardening radius, $a_h=GM_2/4\sigma^2$, and
only few low-angular momentum stars have orbits with 
semi-major axis $\lesssim r_{\rm inf}$. This is not true for unequal-mass binaries, where
$r_{\rm inf} \gg a_h$, and almost all interacting stars are bound to $M_1$.

In this paper we develop a formalism for studying the dynamics of MBHBs 
embedded in gravitationally-bound stellar cusps. The plan is as follows. In \S~2 we describe 
our suite of three-body scattering experiments between the black hole pair and ambient 
bound stars. Our numerical results are 
used in \S~3 and \S~4 to construct a ``hybrid model'' of binary dynamics and investigate the orbital decay 
and shrinking of MBHBs in time-evolving stellar cusps. The properties of the ejected HVSs are 
discussed in \S~5. The massive black hole (Sgr A$^*$) in the Galactic Center and the stars around it
offer a unique opportunity to study stellar dynamics in the extreme environment
around a relativistic potential. The scattering of stars bound to Sgr A$^*$ by an 
inspiraling intermediate-mass black hole (IMBH) is treated in \S~6. 
Finally, we present a brief summary in \S~7.
 
 \section{SCATTERING EXPERIMENTS WITH BOUND STARS} 

Consider a MBH of mass $M_1$ surrounded by a stellar cusp (with density profile 
$\rho\propto r^{-\gamma}$, $\gamma>0$), interacting with a secondary ``intruder" 
hole of mass $M_2<M_1$. As the binary
separation decays, the effectiveness of dynamical friction slowly declines because
distant stars perturb the binary's center of mass but not its semi-major axis.
The bound pair then loses orbital energy by capturing ambient stars and ejecting them 
at much higher velocities, a three-body scattering process known as the `gravitational slingshot'.
For unequal-mass binaries (mass ratio $q\lesssim 0.1$), 
$a_h\ll r_{\rm inf}$, implying that the stellar mass 
inside $a_h$ is $M_{\rm cusp}(<a_h)\ll M_1$. 
In this case, the contribution to the potential energy given by the stellar 
distribution during a binary-star interaction can be ignored, and 
the problem can be tackled by means of three-body scattering experiments.
Following the guidelines fully described in Paper I, we performed numerical 
experiments to study the interaction of a MBHB with a star of mass $m_*$ 
($m_*\ll M_2<M_1$) bound to $M_1$. Note that, for extreme mass ratios $q\ll 1$, 
the encounter is essentially a two-body scattering as the star and the secondary hole 
move in the static potential of the primary. 

\subsection{Initial conditions and orbit integration}\label{sec4.1.2}

The integration of the three-body encounter equations is greatly simplified by 
setting the center of mass of the binary at rest in the origin of the coordinate system.
The binary orbits counterclockwise in the $(x,y)$ plane, and the apoastron of $M_2$
is located along the positive $x$-axis. Stars are drawn from a spherical isotropic 
distribution bound to $M_1$. The problem is completely defined by ten variables:

\begin{enumerate}
\item the binary mass ratio $q\equiv M_2/M_1$;
\item the binary eccentricity $e$;
\item the stellar mass $m_*$;
\item the initial distance between the primary hole and the star, 
$r_*\equiv |{\bf r_*}|$;
\item the specific energy of the stellar orbit around $M_1$, $E_*=-GM_1/(2a_*)$, 
or equivalently the semi-major axis of the stellar orbit, $a_*$;
\item the specific angular momentum of the stellar orbit $L_*=|{\bf v}\times {\bf r_*}|$,
where $v\equiv |{\bf v}|$ is the stellar velocity relative to $M_1$;
\item four angles: $\theta$ and $\phi$ describing the latitude and
longitude of the star, $\psi$ defining the orientation of $v_\perp$ (the stellar 
velocity component normal to ${\bf r_*}$) i.e. selecting the orbital plane of the star, and 
$\Psi$ the initial binary phase.
\end{enumerate}

\begin{figurehere}
\vspace{0.5cm} 
\centerline{\psfig{figure=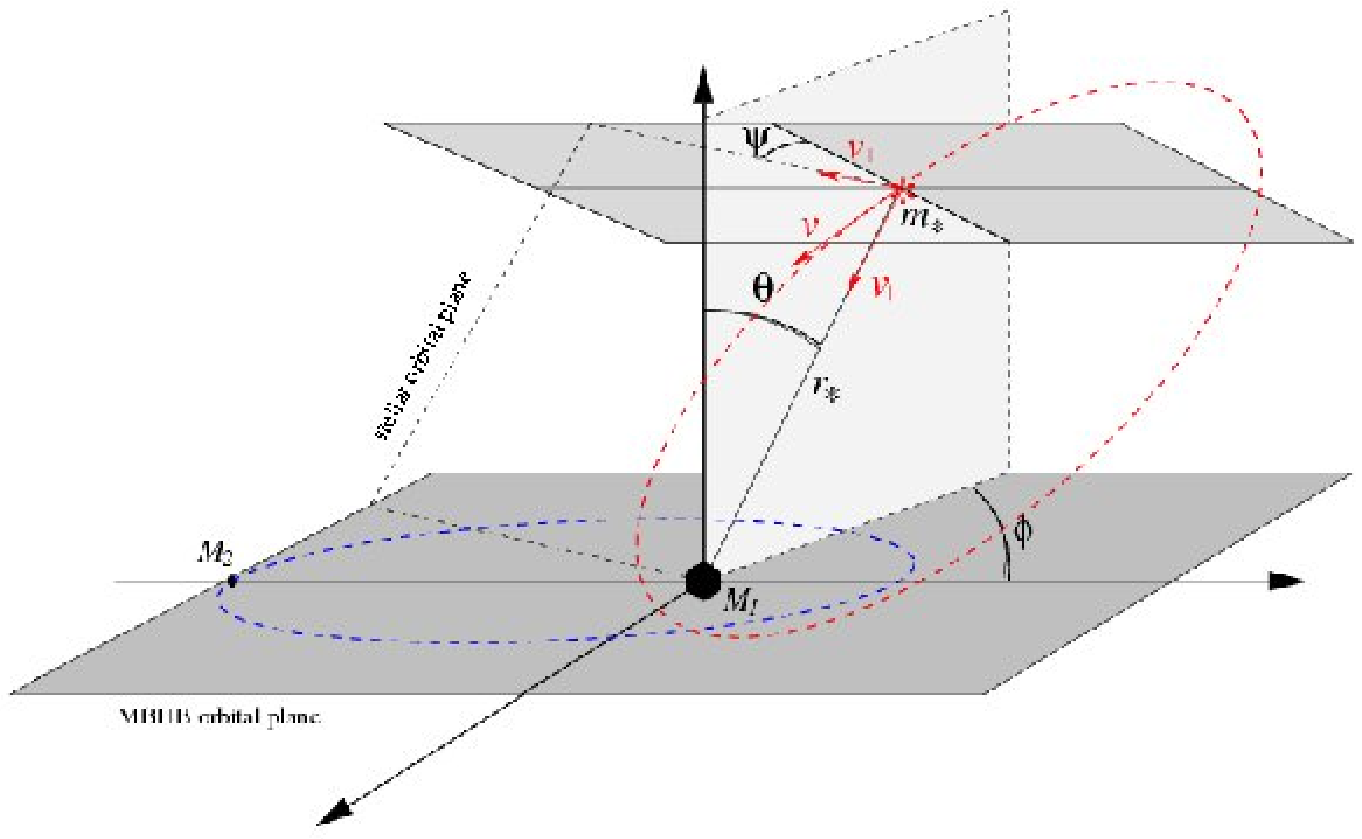,width=3.5in,angle=270}}
\vspace{-0.0cm}
\caption{\footnotesize Geometry of the three-body scattering experiments. 
The stellar orbits is determined by $E_*$ and $L_*$, while its orientation with respect to the 
binary plane is determined by angle $\psi$. See text for definition of all symbols.}
\label{fig:incond}
\vspace{+0.5cm}
\end{figurehere}

Compared to the unbound scattering problem (which is defined by nine variables, see Paper I), 
there is here the extra variable $r_*$, and $E_*$ and $L_*$ replace the asymptotic initial 
speed and impact parameter of the incoming field star.
A sketch of the experiment set-up is given in Figure~\ref{fig:incond}. In each numerical
integration, the binary mass $M=M_1+M_2$ and separation $a$ are set equal to unity, and
the stellar mass to $m_*=10^{-6}M$. Scattering events are simulated in sets of $5\times 10^4$ 
trials for fixed values of $q$ and $e$. We sample five values of the binary eccentricity, 
$e=0, 0.1, 0.3, 0.6, 0.9$, and five values of the binary mass ratio, $q=1/9, 1/27, 1/81, 
1/243, 1/729$, for a grand total of 25 models. The distribution of stars bound to 
$M_1$ is modeled as follows. The stellar semi-major axis $a_*$ is randomly sampled from 
fifty logarithmic bins spanning the range $0.03 a<a_*<10 a$. The angular
momentum is sampled in the interval $[0,L_{*,\rm max}^2]$ according to an equal probability 
distribution in $L_*^2$, where $L_{*,\rm max}^2=GM_1a_*$
is the specific angular momentum of a circular orbit of radius $a_*$. 
A population of stars with such distribution in $L_*^2$ has mean eccentricity 
$\langle e_* \rangle=0.66$, corresponding to $\langle v_\perp \rangle=2\langle 
v_\parallel \rangle$. This condition defines an isotropic stellar distribution 
(e.g. Quinlan, Hernquist, \& Sigurdsson 1995).

The quantities $E_*$ and $L_*$ define the shape of the stellar orbit. 
We sample the initial value of $r_*$ from the distribution ${\cal P}(r_*)dr_*$, 
\begin{equation}
{\cal P}(r_*)dr_*=\frac{2|E_*|^{3/2}}{\pi G M_1\sqrt{E_*-\frac{L_*^2}{2r_*^2}+
\frac{GM_1}{r_*}}}\,\,dr_*,
\end{equation}
which is proportional to the the fraction of time spent by the star at distance between $r_*$
and $r_*+dr_*$ from $M_1$. The probability ${\cal P}(r_*)$ is defined in the range 
$r_{*,-}<r_*<r_{*,+}$, where 
\begin{equation}
\left\{
\begin{array}{ll}
r_{*,-}=L_*^2/(GM_1+\sqrt{G^2M_1^2+2E_* L_*^2})
\\
r_{*,+}=L_*^2/(GM_1-\sqrt{G^2M_1^2+2E_* L_*^2}).
\end{array}\,.
\right.
\end{equation}
The angles $\theta$ and $\phi$ are randomly 
generated to reproduce a uniform density distribution over a spherical 
surface centered on $M_1$, while the orientation angle $\psi$  
is chosen from a uniform distribution in the range $[0,2\pi]$. 
We start numerical integration with $M_2$ at its apoastron ($\Psi=0$), and have checked 
that the chosen initial phase of the binary does not affect our results.
The orbit of the pair is in the $x-y$ plane with the center of mass at coordinates $(0,0,0)$.  

The nine coupled, second order, differential equations of the three-body problem are integrated 
using the most recent version of the subroutine DOPRI5, based on an 
explicit Runge-Kutta method of order 4(5) due
to Dormand \& Prince (1978). A complete description of the integrator
can be  found in Hairer, Norsett, \& Wanner (1993).
The integration is stopped if any of these events occurs:
\begin{itemize}
\item the star leaves the sphere of radius $r=[10^{10}\mu/M]^{1/4}a_0$ with 
positive total energy, where $\mu=M_1M_2/M$ is the reduced mass of the binary
and $a_0$ its initial semi-major axis. At $r$ the force induced
by the quadrupole moment of the binary is 10 orders of magnitude
smaller than the total force acting on the star at a distance $a_0$;

\item the physical integration time exceeds 1 Gyr;

\item the integration reaches $10^8$ time steps. This typically corresponds 
to the complete integration of $10^4-10^5$ binary orbits.
\end{itemize}

\begin{figurehere}
\vspace{0.5cm} 
\centerline{\psfig{figure=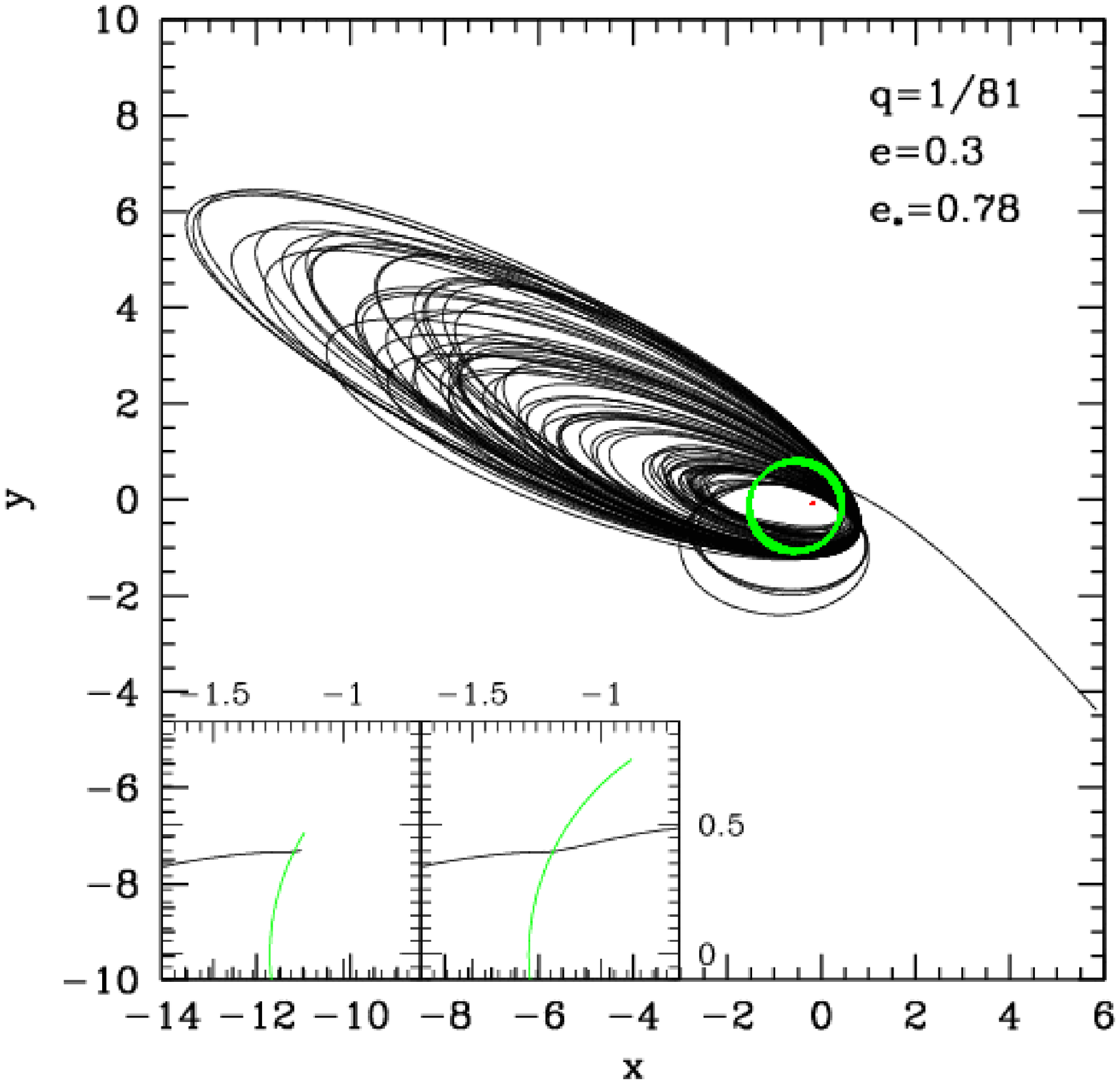,width=3.6in}}
\vspace{-0.0cm}
\caption{\footnotesize Example of a strong three-body interaction unbinding the star. The parameters of the system are 
listed in the figure. The star is ejected towards the lower right with a kick velocity $V\approx 
0.05 V_c$. The thick circle marks the orbit of $M_2$, while the small central dot marks
the location of $M_1$. The insets zoom-in at the moment of the scattering between the 
star (approaching from the left) and $M_2$ (approaching from below).}
\label{fig:boundej}
\vspace{+0.5cm}
\end{figurehere}
\begin{figurehere}
\vspace{0.5cm} 
\centerline{\psfig{figure=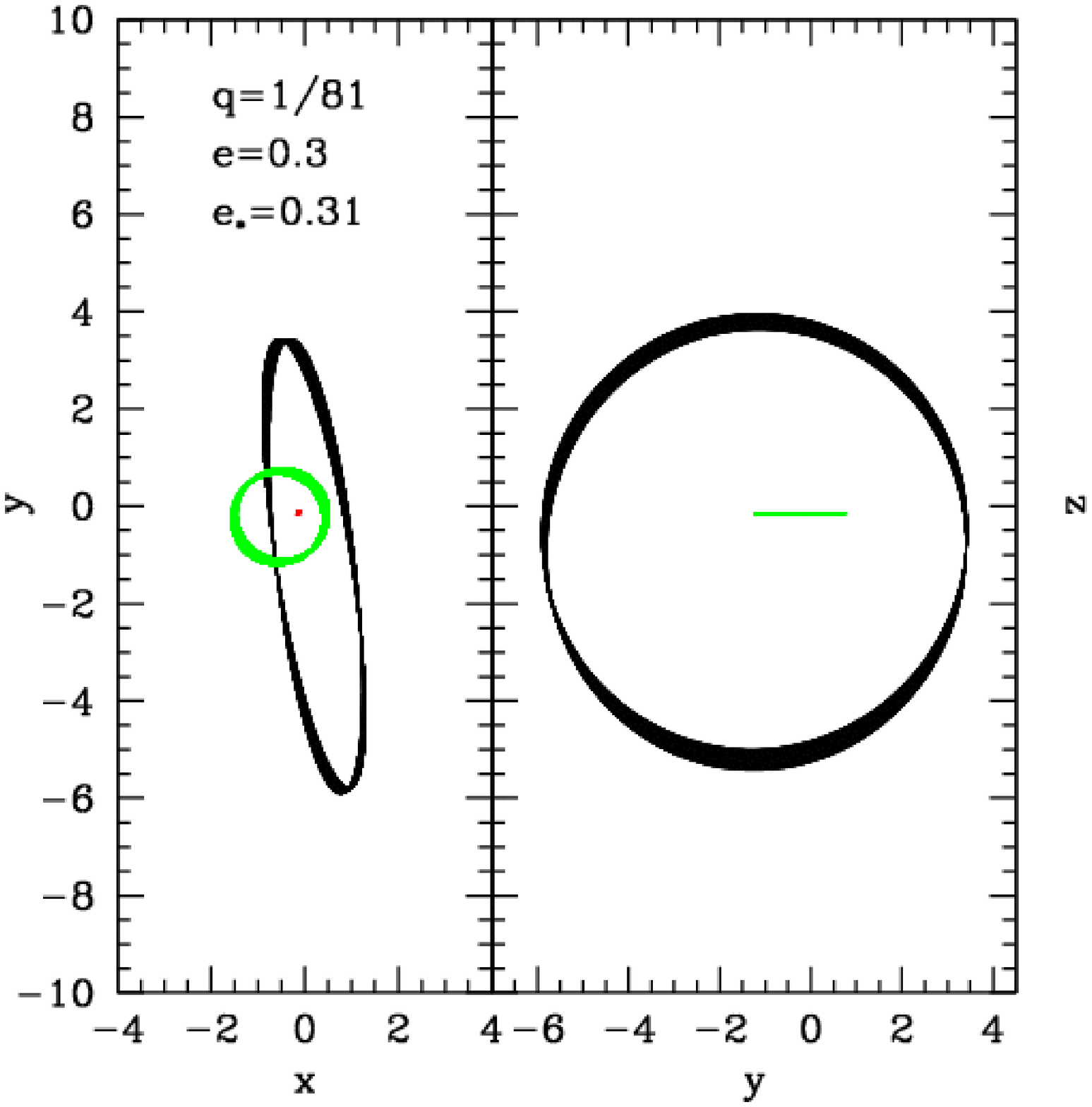,width=3.6in}}
\vspace{-0.0cm}
\caption{\footnotesize Example of a weak three-body interaction. {\it Left panel}: projection onto the 
binary ($x-y$) plane.  {\it Right panel}: projection onto the $y-z$ plane}
\label{fig:boundweak}
\vspace{+0.5cm}
\end{figurehere}

Examples of integrated stellar orbits are given in Figures \ref{fig:boundej} and \ref{fig:boundweak}.
Figure \ref{fig:boundej} shows a strong interaction: 
the star has a close encounter with $M_2$ and leaves the binary
sphere of influence with positive energy. 
Depending on the value of $a_*$ we have a variable number
of numerically ``abandoned events'', i.e., bound triple systems that do not readily produce 
an ejection. Figure \ref{fig:boundweak}
shows one of such events: the periastron of the star
is much larger than the binary orbital separation $a$, and the star is not perturbed 
by the binary quadrupole field, except for the precession of its periastron.

\subsection{Tests}

We have performed a number of tests to check the sensitivity of our results on numerics.
Because of the intrinsically chaotic nature of the three-body problem, the properties of 
the ejected stars are meaningful only for a statistically-significant sample. 
The integration of the full three-body problem allows us to directly control
the conservation of total energy and angular momentum. The code adjusts the
integration stepsize to keep the fractional error per step in position and 
velocity, $\epsilon$, below $10^{-13}$. This allows a total energy conservation 
accuracy of $\Delta E/E \sim 10^{-9}$ in a single orbit integration, i.e. for $m_* /M\simeq 10^{-6}$ 
the energy of the star is conserved at level of one part in a thousand during a single orbit. 
We varied $\epsilon$ between $10^{-11}$ and $10^{-15}$ and $m_*/M$ between $10^{-5}$ and 
$10^{-7}$, and found no significant differences in the statistics of the ejected population. 
We have also checked that the mean energy and angular momentum exchanges scale linearly 
with $m_*$. A set of longer orbit integrations was performed to  
test that the number of abandoned events does not depend on the 
finite numbers of timesteps allowed, again finding no systematic decrease in such a number.

\subsection{Outputs}

Each logarithmic bin in $a_*$ was sampled by $10^3$ stars. We calculated the 
mean energy and angular momentum exchange between the MBHB and the stars, the 
final velocity and angular distribution of the scattered stars, 
the fraction $f_{\rm ej}$ of stars that are ejected in the interaction, 
and the ejection timescales. The fraction $f_{\rm ej}$ is plotted versus $a_*/a$
in Figure \ref{fig4.1} for different values of the binary parameters $q$ and $e$. Eccentric 
binaries can eject stars that are initially very tightly bound to $M_1$, i.e. with 
$a_*$ as small as $0.1a$. For $a_*\gtrsim 3a$, the expelled fraction 
declines significantly, regardless of binary eccentricity. It is also evident that the 
ejection process is more efficient for large values of $q$. 

\begin{figurehere}
\vspace{0.5cm} 
\centerline{\psfig{figure=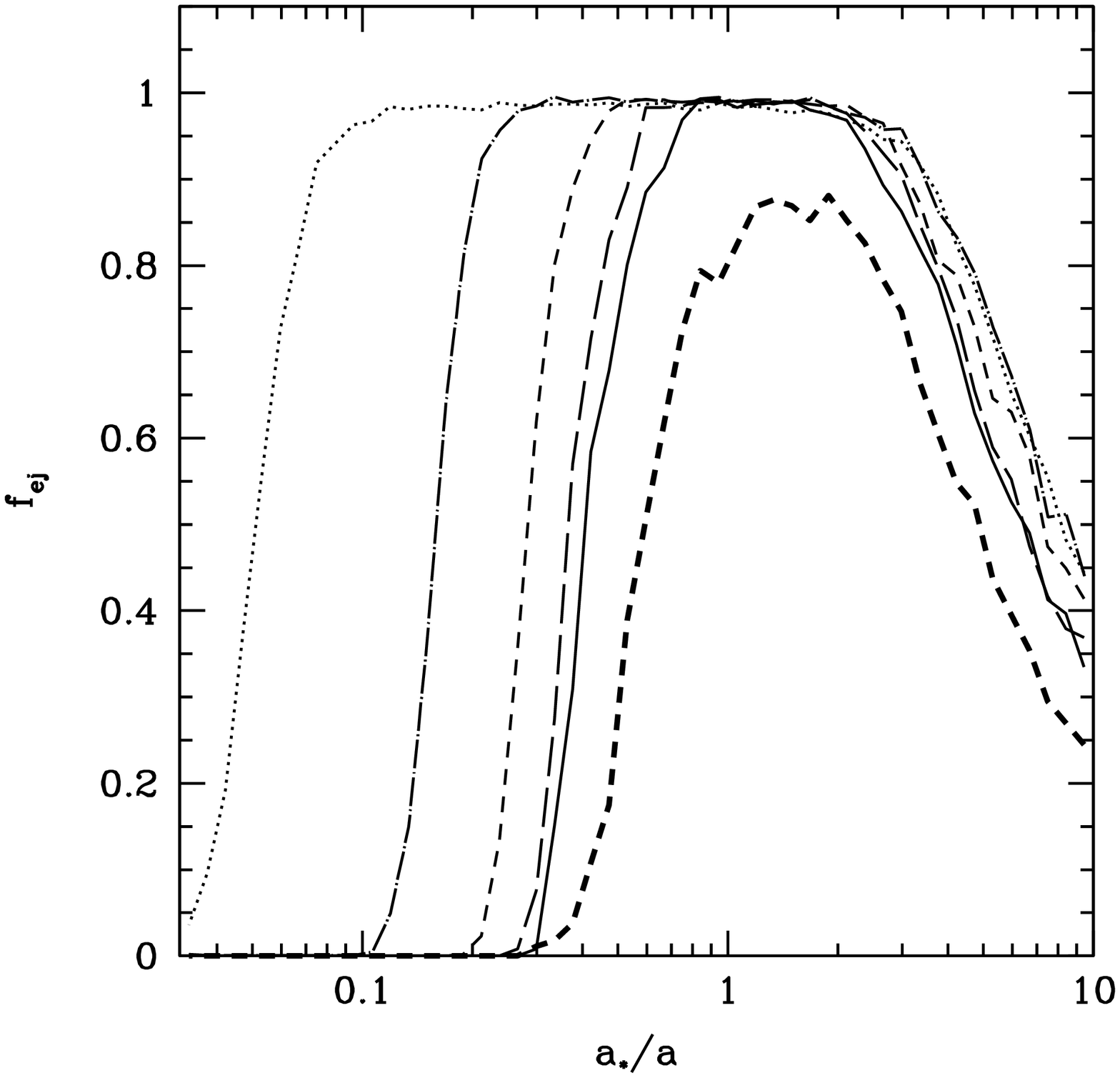,width=3.6in}}
\vspace{-0.0cm}
\caption{\footnotesize Fraction of stars ejected in the interaction as a function of $a_*/a$. Thin curves 
are for $q=1/9$ and different eccentricities: $e=0$ ({\it solid line}), $e=0.1$ 
({\it long-dashed line}), $e=0.3$ ({\it short-dashed line}), $e=0.6$ ({\it dot-dashed line}), 
$e=0.9$ ({\it dotted line}). The thick dashed line is for $q=1/729$ and $e=0.3$.}
\label{fig4.1}
\vspace{+0.5cm}
\end{figurehere}

Figure~\ref{fig4.3} shows the mean fractional eccentricity change of the pair after each
scattering, $\langle \Delta e/e\rangle$, as a function of $a_*/a$. This
quantity is found to scale linearly with $m_*/M$. 
Stars with $a_*<a$ typically tend to reduce the binary eccentricity, while stars with
$a_*>a$ work in the opposite direction.
An eccentric binary spends most of its period near its apocenter, so in the case
$a_*>a$ the probability of a close star-binary encounter (and
subsequent star ejection) is maximal at binary apocenter. In the instantaneous
interaction the binary velocity decreases.  As close to the
apocenter the binary velocity is nearly tangential, the binary is forced on a more radial orbit.
On the contrary, a star with $a_*<a$ ``feels'' the secondary hole $M_2$ when this approaches the primary
$M_1$ at a distance $\sim a_*$. At that point, the interaction with the star is unlikely
to occur close to the pericenter of $M_2$ (being the time spent by $M_2$ at the pericenter 
very small), and typically also extracts a large radial component from the velocity of 
the secondary black hole, hence causing circularization.
\begin{figurehere}
\vspace{0.5cm} 
\centerline{\psfig{figure=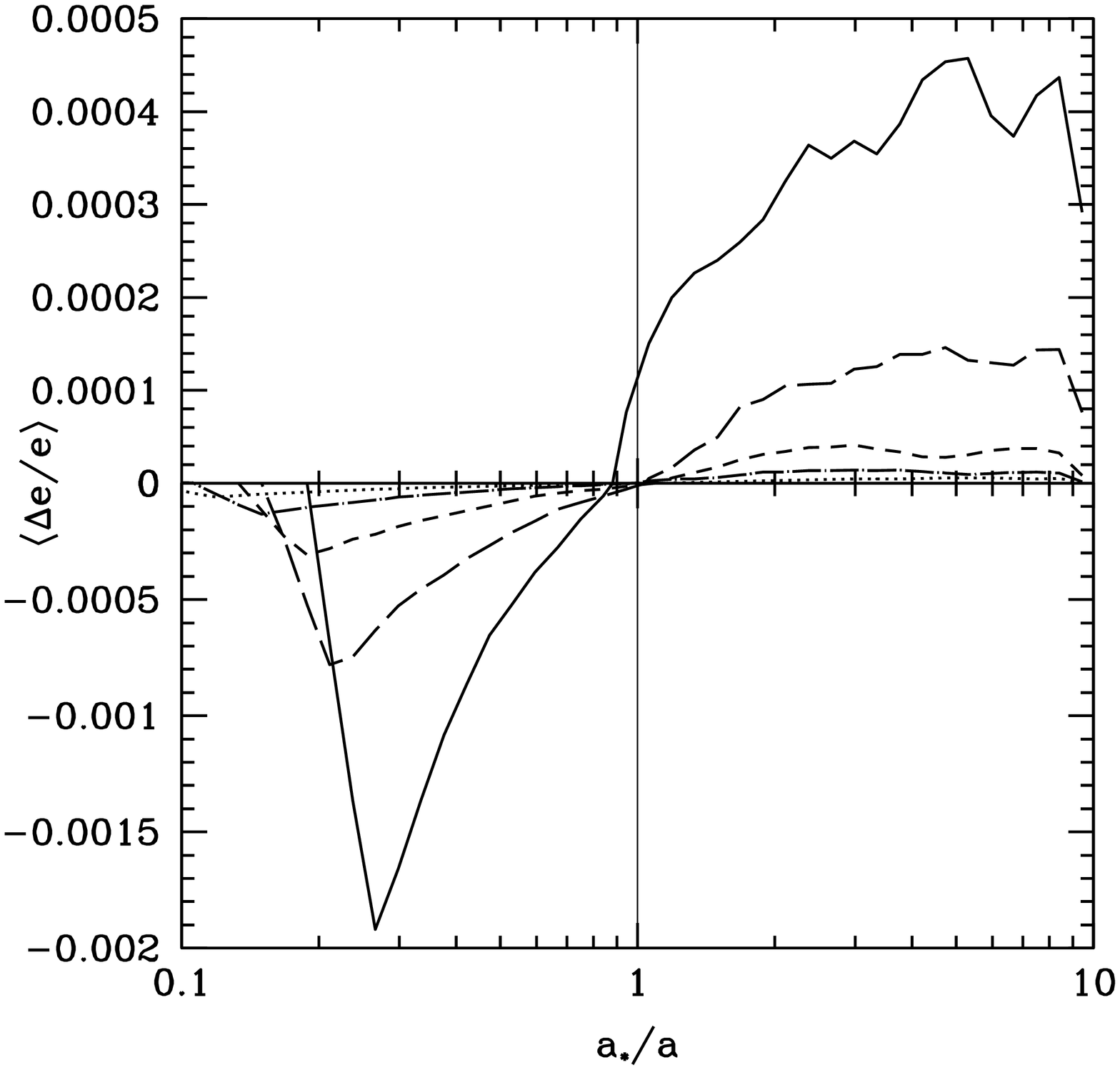,width=3.6in}}
\vspace{-0.0cm}
\caption{\footnotesize Mean fractional eccentricity change, $\langle \Delta {e}/e\rangle$, of the binary
after a scattering, as a function of $a_*/a$. The assumed initial eccentricity is $0.6$, and the different 
curves are for $q=1/729$ ({\it solid line}), $q=1/243$ ({\it long-dashed line}), 
$q=1/81$ ({\it short-dashed line}), $q=1/27$ ({\it dot-dahed line}), and $q=1/9$ ({\it 
dotted line}). Note the different scales of the positive and negative $y$-axis.}  
\label{fig4.3}
\vspace{+0.5cm}
\end{figurehere}
\begin{figurehere}
\vspace{0.5cm} 
\centerline{\psfig{figure=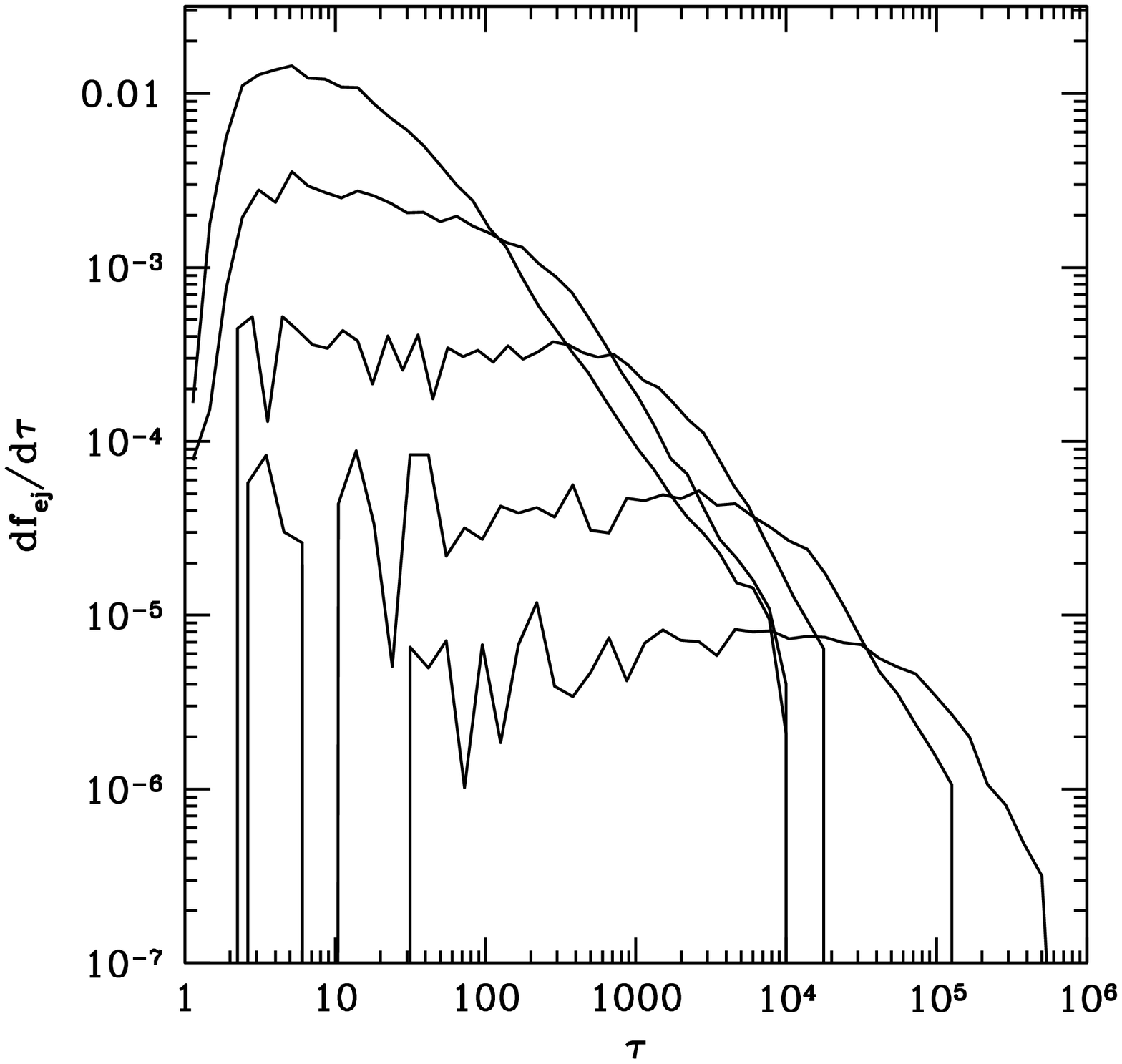,width=3.6in}}
\vspace{-0.0cm}
\caption{\footnotesize Differentail distribution of ejection timescales $\tau$. The assumed initial
eccentricity is $0.3$, and the different curves, from top to bottom, are for $q=1/9, 1/27, 
1/81, 1/243,$ and $1/729$.
} 
\label{fig4.4}
\vspace{+0.5cm}
\end{figurehere}
Note that in this case the absolute value
of $\Delta{e}$ is larger, according to the $a/a_*$ dependence
in equation (\ref{deltaebound}) (see the different positive and negative $y$-axis
scales in Fig.~\ref{fig4.3}). We will return on this point in Section 3, giving a simple,
physically motivated, mathematical derivation of this qualitative argument.

Figure~\ref{fig4.4} depicts the fractional number of stars ejected in the time interval
$\tau$ and $\tau+d\tau$ as a function of the ejection timescale $\tau$. The latter is 
measured in units of the binary orbital period at separation 
$a_0\equiv a(t=0)$, $P_0=2\pi\sqrt{a_0^3/(GM)}$, and 
is defined as the time elapsed from the start of numerical integration to the moment 
the interacting star reaches, with positive energy, a distance $\gg a$ from the 
binary center-of-mass. For large values of $q$, it typically takes the MBHB just a few 
orbits to expel the star, while lowering $q$ makes the distribution of slingshot 
timescales broader and flatter. The ejection rate remains approximately constant, or 
decreases slowly, for $\tau  \lta  5/q^2$, and drops dramatically afterwards. 
The binary eccentricity plays no role in all the tested cases. Figure~\ref{fig6bis} shows 
the mean ejection timescale (i.e., the integral of $\tau (df_{\rm ej}/d\tau)$ in $d\tau$)
as a function of the star eccentricity $e_*$, for different 
values of $q$ and $e$. In first approximation, one can assume 
that the cusp remain isotropic during the binary orbital evolution. 

\begin{figurehere}
\vspace{0.5cm} 
\centerline{\psfig{figure=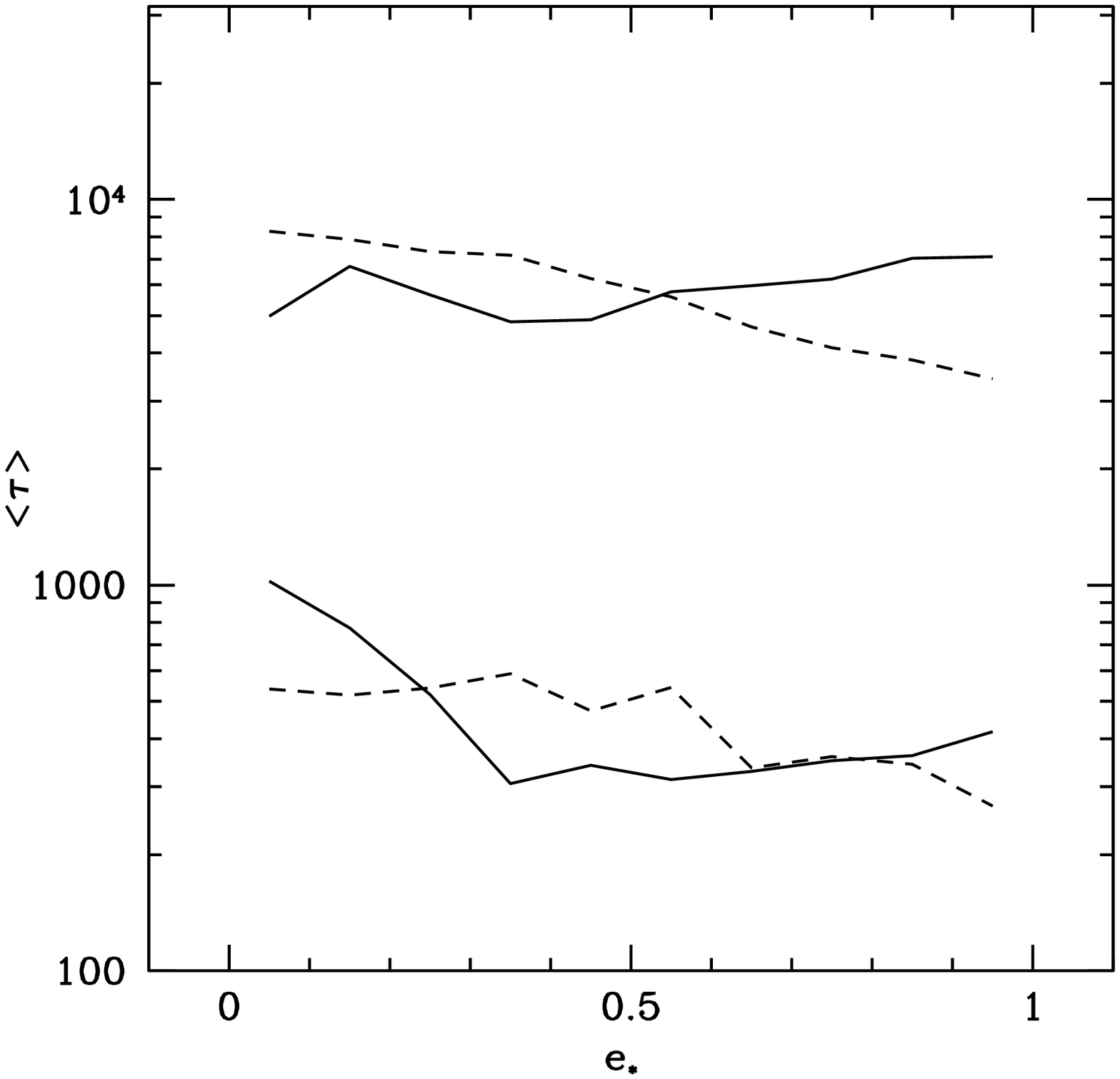,width=3.6in}}
\vspace{-0.0cm}
\caption{\footnotesize Mean ejection timescale $<\tau>$ of stars as a function of the star eccentricity $e_*$, 
assuming a MBHB with $e=0.1$ ({\it solid line}) and $e=0.9$ ({\it dashed lines}). The 
two upper (lower) curves are for $q=1/81$ ($q=1/9$).} 
\label{fig6bis}
\vspace{+0.5cm}
\end{figurehere}

We checked that our results are practically the same considering $5a$ rather 
than $10a$ as the outer boundary in the sampled $a_*$ distribution. This is because 
the binary--stars interaction is dominated, in terms of energy and angular momentum exchange, 
by those stars whose $a_*$ is $\sim a$.

From our scattering experiments, we can finally derive the bivariate distribution 
functions $h_1(V,\theta|a_*)$ and $h_1(V,\phi|a_*)$, along the same line described in Paper II. 
For a given $a_*$, we record the number of of stars with ejection speed in the interval $V,V+dV$, 
leaving the binary with latitude (longitude) at infinity in the interval $\theta,\theta+d\theta$ 
($\phi,\phi+d\phi$). The differential distributions are normalized as follows:
\begin{equation}\label{eq:norm}
\int_0^\infty dV \int_0^\pi d\theta\,h_1(V,\theta|a_*)=\int_0^\infty dV \int_0^{2\pi}\,
d\phi\,h_1(V,\phi|a_*)=1. 
\end{equation}
The subscript ``1" is meant to indicate that the scattering experiments are performed for a binary at 
separation $a=1$.

\section{HYBRID MODEL OF BINARY DYNAMICS}

To study the impact of the gravitational slingshot on the dynamical evolution of a MBHB, 
we have developed a self-consistent hybrid model, in which numerical
results of scattering experiments in a fixed stellar background 
are coupled to an analytical formulation of loss-cone depletion. This technique
allows us to simultaneously follow the orbital decay of the pair as well
as the time evolution of the stellar cusp. The hybrid model is similar, in spirit, to 
the case extensively discussed in Paper II of an unbound stellar population interacting 
with a black hole pair. Scattering experiments are performed for a binary at fixed orbital separation $a$, 
and the results scaled to any value of $a$. Then, by specifying the pace at which $a$ evolves, 
we can solve for the binary orbital decay with time. Mathematical details, however, are 
different from the scheme developed in Paper II. In the unbound case, the pair interacts 
with a stellar population approaching its sphere of influence at a given rate, 
and the typical three-body interaction is a ``fast" process. 
For stars bound to $M_1$ instead, the interacting stellar population is in 
place from the beginning. The temporal evolution of the system is then determined by the 
ejection timescale rather than by supply rate. 

It is convenient to describe the stellar cusp
using $dN_*/da_*$, the differential number of stars orbiting 
$M_1$ with semi-major axis in the range $a_*,\, a_*+da_*$. 
For an isotropic stellar distribution  
\begin{equation}\label{eq:dnda}
\frac{dN_*}{da_*}=C\, 4\pi a_*^2\rho(r=a_*),
\end{equation}
where the stellar density profile is
\begin{equation}\label{eq:density}
\rho(r)=\rho_0 \left(\frac{r}{r_0}\right)^{-\gamma},
\end{equation}
and $C$ is a fudge factor that depends on the cusp slope, 
$C=(1.013, 0.872, 0.831)$ for $\gamma=(2, 1.75, 1.5)$ (Ivanov, Polnarev, \& Saha 2005).
Two differential equations determine the rate of change of orbital separation and eccentricity:
\begin{equation}\label{eq:aev}
\frac{da}{dt}=-\frac{2a^2}{GM_1M_2}\int_0^\infty \Delta {{\cal E}}
\frac{d^2N_{\rm ej}}{da_* dt}da_*,
\end{equation}
and
\begin{equation}\label{eq:eev}
\frac{de}{dt}=\int_0^\infty \Delta {e}\frac{d^2N_{\rm ej}}{da_* dt}da_*. 
\end{equation}
We start numerical integration at time $t=0$, orbital separation $a_0$, and binary eccentricity $e_0$. 
The terms $\Delta e$, $\Delta {\cal E}$, and $d^2N_{\rm ej}/da_* dt$ are measured from our 
scattering experiments. 

The scaling of the eccentricity change with binary parameters 
$m_*/M,e,q,$ and $a/a_*$ can be derived from (Quinlan 1996) 
\begin{equation}
\Delta e=-\frac{(1-e^2)}{2e}\left(\frac{\Delta {\cal E}}{{\cal E}}+
\frac{2\Delta {\cal L}_z}{{\cal L}_z}\right),
\end{equation}
where ${\cal L}_z=\mu\sqrt{GMa(1-e^2)}$ and ${\cal E}=-GM_1M_2/(2a)$ are the total
angular momentum and energy of the binary, respectively. 
From $\Delta {\cal E}=-\Delta {\cal E}_* \sim -GM_1m_*/(2a_*)$
(in a typical encounter, on average, the star gets a kick $\sim V_c$),
and 
$\Delta {\cal L}_z=-\Delta {\cal L}_* \sim -{\cal L}_*=-m_*\sqrt{GMa_*(1-e^2_*)}$, 
 assuming $M_1\simeq M$, we have  
\begin{equation}\label{deltaebound}
\Delta e \frac{M}{m_*}\,\approx \, \frac{(1-e^2)}{2eq}
\left(\sqrt{\frac{(1-e^2_*)a_*}{(1-e^2)a}}-\frac{a}{a_*}\right).
\end{equation}
This confirms the qualitative argument developed in
Section 2.3 and accounts for the features shown in Figure 5: the mean $\Delta e$ scales 
with the inverse of the mass ratio, and, from $<e_*>\simeq 0.67$, 
results negative for $a_{\star}/a \lta (1-e^2)^{1/3}$ (promoting circularization), and positive for larger 
$a_*/a$ (hence increasing the binary eccentricity).

The binary energy change per scattering, $\Delta {\cal E}$, is related 
to the change of stellar energy $\Delta {\cal E}_*$ by 
$\Delta {\cal E}_* \simeq -\Delta {\cal E}$,
where we have neglected the energy change of the binary center of mass (which is 
a factor $m_*/M$ smaller than $\Delta {\cal E}$). The quantity $\Delta {\cal E}_*$ 
can be written as    
\begin{equation}\label{eq:deltaes}
\Delta {\cal E}_*(a_*,a,e)=
\frac{GM_1m_{*}}{2a_*}+\frac{1}{2}
k\, m_{*}V_c^2(a),
\end{equation}
where $V_c(a)=\sqrt{GM/a}$ is the circular velocity of the binary. The numerical factor $k$ depends 
upon the ratio $a_*/a$, and $e$, and is derived from our numerical experiments. 

The term $d^2N_{\rm ej}/da_* dt$ quantifies the number
of stars orbiting $M_1$ within $a_*$ and $a_*+da_*$ that are ejected from the system in the
time interval between $t$ and $t+dt$. This term depends on $a_*$, $a$, 
$e$ and $t$, and is determined as follows. 
From scattering experiments we derive the distribution $P(\tau,s|e)d\tau$ describing the 
probability that a star at $s\equiv a_*/a$ becomes unbound from a binary with 
eccentricity $e$ in the time interval $(\tau,\,\tau+d\tau)$. Note that the distribution of 
ejection timescales plotted in Figure~\ref{fig4.4} is simply the probability function 
$P$ averaged over $s$, 
\begin{equation} \label{eq:fejp}
\frac{df_{\rm ej}}{d\tau}\,=\, \frac{\int{P(\tau,s|e)ds}}{\int{ds}}.
\end{equation}
When $s \ll1$ (the exact value depending on $q$ and $e$) most stars remain bound to $M_1$, and $P=0$ 
(see Fig. ~\ref{fig4.1}). As $s \rightarrow 1$ all interacting stars are instead expelled. 
We can see from Figure~\ref{fig4.4} that, in the ejection regime, the distribution $P$ is 
nearly independent on $s$ and has the same functional form (as a function of $\tau$) of $df_{\rm ej}/d\tau$.
From $P$ we can then compute $d^2N_{\rm ej}/da_* dt$ as a function of $t$ if we set 
$\tau=(a_0/a)^{3/2}\,t$, where the term $(a_0/a)^{3/2}$ accounts for the change in the time units 
of $P$ as $a$ changes. Though not formally correct, the 
scheme catches the basic physics of the interaction in a time-dependent fashion.
 
Suppose now that in a small time interval $\Delta t_0$ after the beginning of the 
interaction the binary remains at constant separation $a_0$ corresponding to $s_0\equiv a_*/a_0$. 
The number of stars (with semi-major axis in the interval $a_*,a_*+da_*$) 
ejected in such time interval is
\begin{equation}
{\cal F}_0\,\equiv \,\frac{d^2N_{\rm ej}}{da_* dt}(0)\, \Delta t_0\,=\, \frac{dN_*}{da_*}P(0,s_0)\,\Delta t_0,
\label{eq:F0}
\end{equation}
where for simplicity we have omitted the dependence of $P$ on the eccentricity. 
After $\Delta t_0$ and for an interval $\Delta t_1$, the binary settles to a new separation 
$a_1$, corresponding to $s_1\equiv a_*/a_1$. The number of stars ejected in the time interval 
$\Delta t_0,\Delta t_0+ \Delta t_1$ is
\begin{equation}
{\cal F}_1\,=\,\left(\frac{dN_*}{da_*}-{\cal F}_0 \right) \, \frac{P(\Delta t_0,s_1)}
{\int_{\Delta t_0}^\infty{P(t',s_1)dt'}}\,
\Delta t_1\,\left(\frac{a_0}{a_1}\right)^{3/2}. 
\label{eq:F1}
\end{equation}
The integral at the denominator of the right-hand side renormalizes the distribution $P$ in 
the time interval $[\Delta t_0,\,\infty]$ so that the correct number of stars is involved. 
Iterating we have 
\begin{equation}
{\cal F}_j \,=\,\left(\frac{dN_*}{da_*}-\sum_{i=0}^{j-1}{\cal F}_i \right) \, 
\frac{P(t,s_j)}{\int_t^\infty{P(t',s_j)dt'}}\,
\Delta t_j\,\left(\frac{a_0}{a_j}\right)^{3/2},
\label{eq:Fj}
\end{equation}
where $t\equiv \sum_{i=0}^{j-1}\Delta t_i$.  In differential form:  
\begin{equation}
\frac{d^2N_{\rm ej}}{da_* dt}(t)\,=\,\left(\frac{dN_*}{da_*}-\int_0^t{\frac{d^2N_{\rm 
ej}}{da_* dt'}dt'} \right)\,\frac{P(t,s)}{\int_t^\infty{P(t',s)dt'}}\,
\,\left[\frac{a_0}{a(t)}\right]^{3/2}, 
\label{eq:d2Ndiff}
\end{equation}
where $s=s(t)=a_*/a(t)$. 
Simultaneous numerical integration of the tree coupled equations (\ref{eq:aev}), (\ref{eq:eev})
and (\ref{eq:d2Ndiff}) self-consistently solves for the evolution of the binary and the
depletion of the stellar cusp. The integration is performed using the subroutine DOPRI5. 
The time--step is adapted in order to keep the fractional error per step $\leq 10^{-10}$, a value 
much lower than the error associated to the employed linear interpolation of $\Delta e$ and 
$\Delta {\cal E}$. 

The bivariate distributions $h_1(V,\theta|a_*)$ and $h_1(V,\phi|a_*)$ derived  
from our suite of scattering experiments (see \S~2) can be convolved with the ejection 
rate ${d^2N_{\rm ej}}/{da_* dt}$ to compute the final velocity distributions $h(V,\theta)$ and 
$h(V,\phi)$. The procedure is similar to that described in the Appendix of Paper II. 
As the binary shrinks to separation $a<1$, the normalized distribution of ejection velocities for 
stars with semi-major axis in the interval $a_*,\, a_*+da_*$ is 
\begin{equation}\label{eq:fvbin}
h_a(V,\theta|a_*) = \frac{1}{\sqrt{a}}\,h_1\left(\frac{V}{\sqrt{a}},\theta | a_*\right),  
\end{equation}
where the ejection speed $V$ was shifted by the factor $1/\sqrt{a}$ to account for the increase 
in the circular velocity of the binary $V_c$ as the pair shrinks. The prefactor $1/\sqrt{a}$ 
normalizes the distribution according to equation (\ref{eq:norm}). The kick velocity function
of the expelled population as a whole can then be written as 
\begin{equation}\label{eq:fvej}
h(V,\theta)=\frac{\int_1^{a_f}\int_0^\infty\frac{d^2N_{\rm ej}}{da_* da}h_a(V,\theta|a_*)
da_* da}{\int_1^{a_f}\int_0^\infty\int_0^\pi\int_0^\infty\frac{d^2N_{\rm ej}}
{da_* da}h_a(V,\theta|a_*)da_* da d\theta dV},
\end{equation}
where $a_f$ is the final binary separation,  
\begin{equation}\label{eq:dmej}
\frac{d^2N_{\rm ej}}{da_* da}\equiv \frac{d^2N_{\rm ej}}{da_* dt}\frac{dt}{da},
\end{equation}
and $dt/da$ is given by equation (\ref{eq:aev}). The distribution
$h_1(V,\theta|a_*)$ is evaluated for different values of the eccentricity $e$. 
We can account then for the evolution of binary eccentricity by interpolating the $h_1$
distribution on a grid of $e$-values as the orbit decays.  
The above procedure returns the velocity distribution in units
of $V_c(a_0)$. The calculation of $h(V,\phi)$ can be performed following the same lines.

Finally, integrating the rate ${d^2N_{\rm ej}}/{da_* dt}$ over the entire evolutionary history 
of the binary yields the quantity ${dN_{\rm ej}}/{da_*}$, which can be 
compared to ${dN/}{da_*}$ to study the depletion of the stellar cusp by three-body interactions.
The Monte Carlo technique developed to address this point is described in \S~5.1.
The functions ${dN_{\rm ej}}/{da_*}$ and $h(V,\theta)$ can be used to 
check energy conservation, by simply  equating the total energy gained by the stars to the increment
in the binding energy of the black hole pair, i.e.
\begin{equation}
\begin{array}{ll}
m_*\int_0^\infty \frac{dN_{\rm ej}}{da_*}\,\Phi(a_*)\,da_*+

\\
\,\,\,\,\,\,\,\,\,+\frac{1}{2}m_*\int_0^\infty V^2 N_{\rm ej}\,h(V)\,dV=\frac{G M_1 M_2}{2}\left(\frac{1}{a_f}-\frac{1}{a_0}\right),
\end{array}\,
\label{eq:encons}
\end{equation}
where
$\Phi(a_*)=GM_1/a_*$, $h(V)=\int h(V,\theta)d\theta$, and $N_{\rm ej}=\int (dN_{\rm ej}/da_*)\,da_*$. 
We have checked that energy is conserved to better than $\sim 1\%$.

It is important to remark at this point that equations (\ref{eq:aev}) and (\ref{eq:eev}) 
are independent of the absolute value of $M$ and $m_{*}$. Indeed, in equation (\ref{eq:aev}), 
the total number of 
interacting stars is  $\propto M_2/m_{*}$, and the energy exchange term is 
$\propto Mm_{*}$. 
The scaling factor $1/M_1M_2$ cancels out, so that the orbital evolution  
depends only on $q$ and $e$. The same consideration holds for equation (\ref{eq:eev}): 
as $\Delta e \propto m_{*}/M$ while the number
of interacting stars $\propto M_2/m_{*}$, the dependence 
on $M$ and $m_{*}$ cancels out.

\section{BINARY EVOLUTION}

For comparison with our previous results on the scattering of unbound stars (Paper II), it 
is convenient to introduce the stellar velocity dispersion $\sigma$, and model 
the {\it outer} stellar component as a singular isothermal sphere (SIS) with density profile  
\begin{equation}\label{eq:sis}
\rho(r)=\frac{\sigma^2}{2\pi G r^2}.
\end{equation}
We assume that this profile extends inward up to the characteristic radius $r_0$  within
which the total stellar mass is $2M_1$ (Merritt 2004). Matching the inner cusp described by 
equation (\ref{eq:density}) to the outer SIS at $r_0$ yields  
\begin{equation}\label{eq:rbrake}
r_0=(3-\gamma)\frac{GM_1}{\sigma^2}.
\end{equation}
We also assume that stellar-binary interactions start at separation $a_0$ where
the enclosed mass stellar is $M_*(<a_0)=2M_2$, yielding 
\begin{equation}\label{eq:a0}
a_0\,=\,\left(\frac{q}{1+q}\right)^{1/(3-\gamma)}\left[(3-\gamma)\frac{GM}{\sigma^2}\right].
\end{equation}
This assumption is motivated by recent N-body simulations of the hardening of unequal MBHB in 
stellar cusps. Matsubayashi, Makino, \& Ebisuzaki (2007) found that dynamical friction 
is efficient in driving orbital decay only as long as the stellar mass inside the binary semi-major 
axis is $\gta M_2$. Beyond this point, the evolution of the pair is driven by three-body 
interactions with individual stars.   


\subsection{Orbital evolution and eccentricity growth}

Our results on binary orbital decay are summarized in Table 1. A 
MBHB can shrink by factors ranging from 6 to 18 depending on 
$q$, $e_0$, and $\gamma$. The decay factor $a_0/a_f$ grows with $e_0$,
as more eccentric binaries can expel stars that are more tightly bound
(see Fig. \ref{fig4.1}). The importance of the initial eccentricity $e_0$ is 
modest for low values of $\gamma$, since in shallow cusps the number of tightly bound 
stars is small anyway. 
The factor $a_0/a_f$ is also a weakly increasing function of $q$, as 
for higher $q$ the radius of influence of the binary is larger 
(in terms of $a_*/a$, see Fig. \ref{fig4.1}). 
The dependence of the decay factor on $\gamma$ simply reflects the fact that, for steeper
stellar cusps, the mean binding energy of the stars is larger.
\footnote{Note that we have not included the effect of ``returning stars'' on the 
decay of the binary. These are ejected stars that do not escape the
host bulge, return on small impact parameter orbits, and can have a secondary 
super-elastic interaction with the MBHB. The role of secondary slingshots was analyzed
in Paper II, where it was found that they can boost orbital decay by as much as a factor 
of 2 for nearly equal-mass binaries, but do not contribute significantly (less than
$20\%$ for $q=1/9$ and less than $5\%$ for $q=1/81$, see Table 1 of Paper II) to binary 
hardening for very unequal-mass pairs. In the bound case, the effect of returning 
stars is likely to be even smaller. Consider for example the case of an isothermal 
stellar profile: the impact of secondary slingshots is proportional to the number of 
stars that can interact with the MBHB more than once, i.e. to the size of the loss-cone 
after the first interaction. For bound stars this is at least a factor of $\sim 2$
smaller than in the unbound case: so even for $q=1/9$, returning stars would cause 
at most a $\sim 10\%$ increase in binary hardening.}

\begin{tablehere}\label{tab2}
\begin{center}
\begin{tabular}{|cc|cc|cc|cc|}
\tableline
&&\multicolumn{2}{c|}{$\gamma=1.5$}&\multicolumn{2}{c|}{$\gamma=1.75$}&\multicolumn{2}{c|}{$\gamma=2$}\\
\tableline
$q$ & $e_0$ & $a_0/a_f$ & $e_f$ & $a_0/a_f$ & $e_f$ & $a_0/a_f$ & $e_f$\\
\tableline\tableline
   &         0.1&    9.63& 0.608& 9.89& 0.350&   11.38& 0.179\\
   $\frac{1}{9}$ &      0.5&    9.99& 0.972& 11.55& 0.907& 15.77& 0.753\\
   &         0.9&    10.06& 0.998& 11.80& 0.992& 17.73& 0.969\\
\tableline
   &         0.1&    8.06& 0.691& 8.35& 0.532& 10.19& 0.408\\
   $\frac{1}{27}$ &     0.5&    8.26& 0.959& 9.35& 0.862& 12.35& 0.710\\
   &         0.9&    8.27& 0.996& 9.64& 0.988& 14.03& 0.958\\
\tableline
   &         0.1&    6.99& 0.755& 7.75& 0.650& 9.39& 0.542\\
   $\frac{1}{81}$ &     0.5&    6.90& 0.922& 7.81& 0.828& 10.14& 0.717\\
   &         0.9&    6.89& 0.996& 7.81& 0.974& 11.00& 0.937\\
\tableline
   &          0.1&    6.49& 0.906& 7.28& 0.805& 9.30& 0.688\\
   $\frac{1}{243}$ &     0.5&    6.39& 0.971& 7.25& 0.914& 9.92& 0.818\\
   &          0.9&    6.38& 0.962& 7.19& 0.986& 10.09& 0.955\\
\tableline
   &          0.1&    6.12& 0.881& 6.91& 0.814& 8.94& 0.724\\
   $\frac{1}{729}$ &     0.5&    5.95& 0.919& 6.91& 0.869& 9.09& 0.797\\
   &          0.9&    5.94& 0.977& 6.92& 0.953& 9.41& 0.900\\
\tableline
\end{tabular}
\end{center}
\caption{\footnotesize Binary shrinking factors and final eccentricities from the hybrid model.}
\label{tab2}
\end{tablehere}

Binary eccentricity also grows as a function of $q$ and $\gamma$.
A shallow cusp increases the relative importance of stars with large 
$a_*$, and the eccentricity growth is then larger. Moreover, binaries with large 
$q$ are more effective in ejecting stars with $a_*/a\ll1$. 
As these stars act to circularize the binary orbit, nearly equal-mass binaries 
decay following less eccentric orbits. Figure \ref{fig5.1} shows examples of binary 
evolution as a function of time. Note the different
scale of the time axis for the three mass ratios: as shown in Figure~\ref{fig4.4}, 
it takes nearly equal-mass pairs a smaller number of binary orbits to unbind the stellar cusp.

Results for the $\gamma=2$ cusp can also be compared to those obtained in Paper II for the case 
of a binary interacting with a population of unbound stars. (In Paper II, shrinking factors 
were normalized to the ``hardening radius" $a_h\equiv GM_2/4\sigma^2$, where, by definition, 
$a_0=4a_h$ for $\gamma=2$.) Binaries with $e_0=0.1$ and $q=(1/9,1/27,1/81,1/243)$, for example, 
shrink by the factors $a_h/a_f=(2.85,2.55,2.35,2.33)$ according to Table 1, compared 
to the corresponding $a_h/a_f=(2.09,1.49,1.19,1.06)$ for the unbound case, i.e., very 
unequal-mass binaries can only decay by extracting the cusp binding energy. 

\begin{figurehere}
\vspace{0.5cm} 
\centerline{\psfig{figure=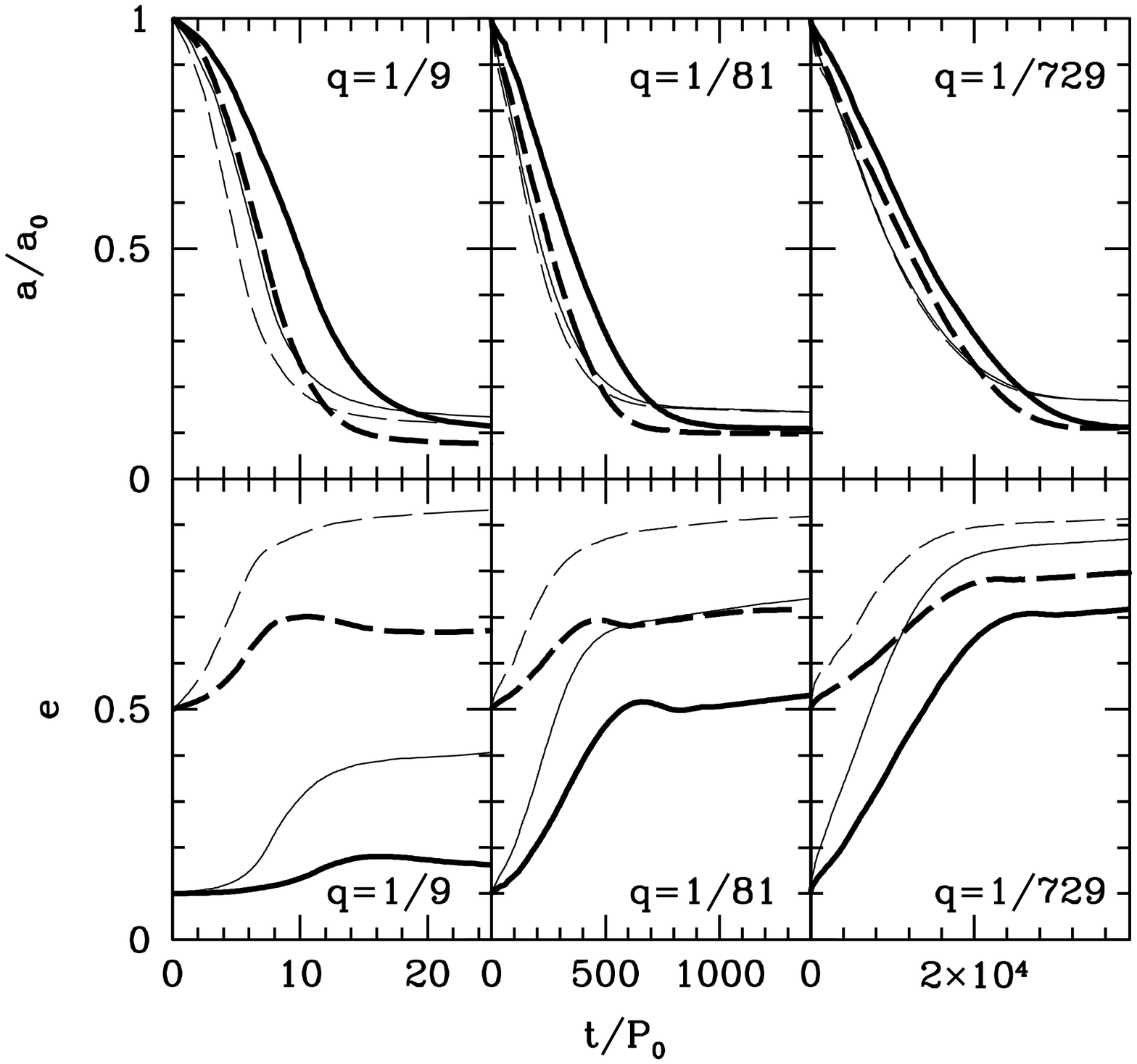,width=3.6in}}
\vspace{-0.0cm}
\caption{\footnotesize Time evolution of the binary semi-major axis ({\it upper panels}) and eccentricity 
({\it lower panels}). {\it Left panels:} $q=1/9$.  {\it Central panels:} $q=1/81$.
{\it Right panels:} $q=1/729$. In all panels the thick lines are for a $\gamma=2$ stellar cusp,
while the thin lines are for $\gamma=1.5$. The initial binary eccentricity is assumed to
be $0.1$ ({\it solid} lines) and 0.5 ({\it dashed lines}).} 
\label{fig5.1}
\vspace{+0.5cm}
\end{figurehere}

The evolution of the eccentricity in the bound and unbound cases is 
compared in Figure~\ref{fig5.1bis}, for two different initial values of $e_0$. Compared to 
the scattering of unbound stars, the binary eccentricity grows to larger 
values. This can be understood by the following argument. In the bound case 
treated here, there is a great deal of stars able to extract angular momentum 
from the binary right from the start. Then, for large $a$, $e$ shows a steep 
increase. When the orbit shrinks, however, stars with $a_*<a$ (increasing 
circularization rather than eccentricity, see Fig.~\ref{fig4.3}) become more and more 
relevant for the binary evolution as stars with $a_*>a$ get progressively depleted, 
and, as result, $e$ increases at a reduced pace. In the unbound case, the 
binary interacts with the same distribution of stars independently upon its 
separation. As, in general, significant changes of $e$ happen whenever a strong 
star-binary interaction occurs, the eccentricity grows more rapidly for small $a$ 
because stars are slower (in terms of binary circular velocity), and hence more 
easily captured in ``quasi" bound orbits before ejection. As a matter of fact, 
the eccentricity growth rate $K$ (see Paper I) increases for $a\ll a_h$. 
In this separation regime, most of the stars interacting with the binary are captured 
in temporarily bound orbits, and would experience a dynamical interaction similar 
to that of bound stars.

\begin{figurehere}
\vspace{0.5cm} 
\centerline{\psfig{figure=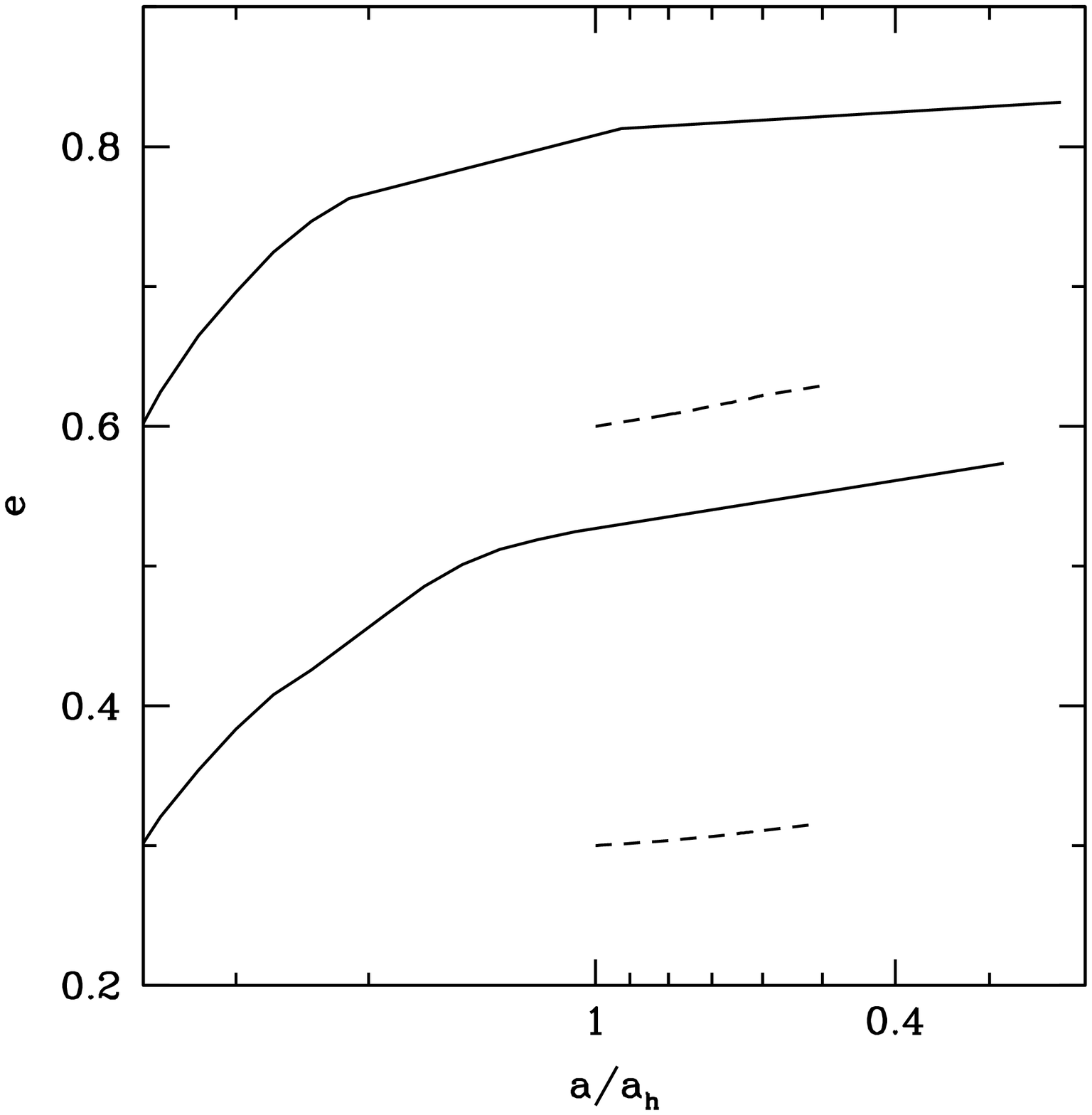,width=3.6in}}
\vspace{-0.0cm}
\caption{\footnotesize Evolution of binary eccentricity as a function of orbital separation, for the case of
interactions with unbound stars (Paper I II, {\it dashed lines}), and with a bound cusp 
({\it solid lines}). The binary is embedded in an SIS, its mass ratio is $q=1/9$, 
and its initial eccentricity is $e_0=0.3$ and 0.6. Note that the scattering experiments 
with unbound stars start at $a=a_h$, while those with bound stars start at $a=4a_h$ 
(see text for details). In the unbound case the evolution of eccentricity 
is negligible ($K\approx 0$, Sesana et al. 2006) for $a>a_h$. 
} 
\label{fig5.1bis}
\vspace{+0.5cm}
\end{figurehere}

\subsection{Final coalescence}\label{s5.4.2}

Consider now, as in Paper II, the separation at which a MBHB can coalesce in less than 
1 Gyr because of the emission of gravitational waves (GWs):
\begin{equation}\label{eq:agwbound}
a_{\rm GW}\simeq \frac{a_h}{250}\left(\frac{1+q}{q}\right)^{3/4}
M_{1,6}^{1/4}F(e)^{1/4},
\end{equation}
where $M_{1,6}\equiv M_1/10^6\,\msun$ and,    
to 4th order in $e$,
\begin{equation}
F(e)\equiv (1-e^2)^{-7/2}\left(1+\frac{73}{24}e^2+\frac{37}{96}e^4\right)
\label{eq:ecce}
\end{equation}
(Peters 1964). Assuming as before that three-body interactions begin operating  
at separation $a_0$, and combining equations (\ref{eq:a0}) and (\ref{eq:agwbound}), 
we get
\begin{equation}\label{eq:afsuagwb}
\frac{a_f}{a_{\rm GW}}\simeq10^3\,\frac{a_f}{a_0}\,
(3-\gamma)\left(\frac{q}{1+q}\right)^{(1+\gamma)/(12-4\gamma)}\,M_{1,6}^{-1/4}F(e)^{-1/4}. 
\end{equation}
If $a_f/a_{\rm GW}<1$, then a MBHB would coalesce in less than 1 Gyr after 
interaction with a bound stellar cusp. As shown in the ``coalescence diagram" of Figure \ref{fig21}, 
eccentric, massive, very unequal-mass binaries embedded in steep stellar cusps are 
favored to reach coalescence. The steeper the cusp, the wider the portion of the $M_1-q$ 
plane where coalescence can be reached within 1 Gyr. 
 
\begin{figurehere}
\vspace{0.5cm} 
\centerline{\psfig{figure=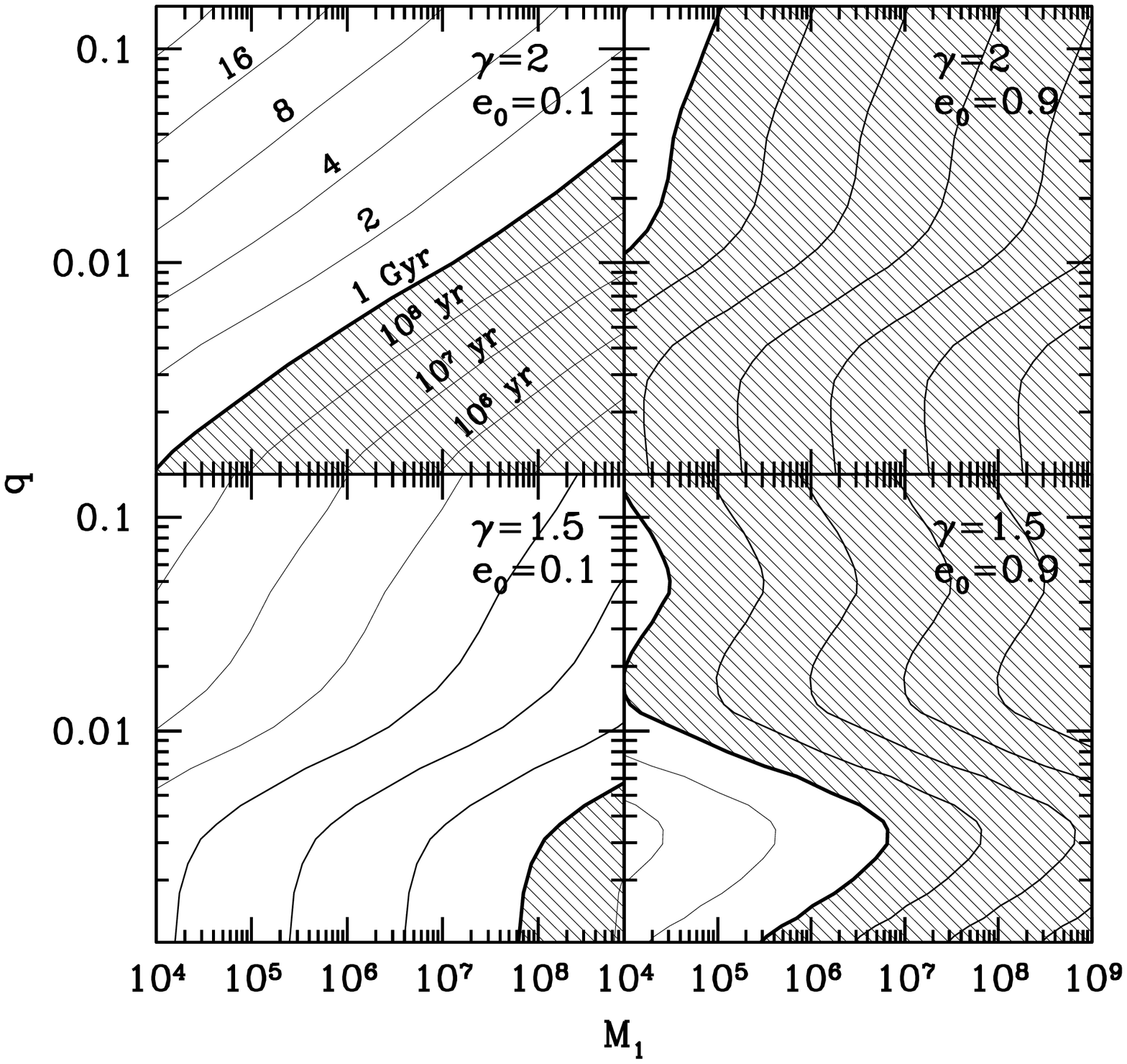,width=3.6in}}
\vspace{-0.0cm}
\caption{\footnotesize Coalescence diagram. In each panel, the {\it thick line} delimits those MBHBs that 
can coalesce, because of GW emission, in less than 1 Gyr starting from separation 
$a_f$ ({\it shaded area}). As shown in the upper left panel, level curves to the left of 
such line are labelled according to the ratio $a_f/a_{\rm GW}$ reached by the binary at the end
of the shrinking process (equation \ref{eq:afsuagwb}), while level curves to the right 
are labelled according to the the coalescence time at $a_f$.} 
\label{fig21}
\vspace{+0.5cm}
\end{figurehere}

\section{EVOLUTION OF THE STELLAR CUSP}

\subsection{Cusp erosion}\label{sec5.3.2}

The orbital decay of the pair occurs following the ejection of a mass $\sim 2-4\,M_2$ 
(depending on cusp slope and binary eccentricity) and results in the progressive erosion of 
the stellar cusp. Eccentric binaries can shrink rapidly by scattering at pericenter a fewer number
of deeply bound stars. More mass is expelled in the case of shallower cusps, because of the 
larger number of stars surrounding the binary just outside $a_0$. 
Our hybrid approach allows us to 
compute the binary--driven evolution of the stellar density profile, once isotropy is assumed. 
At this stage, we do not perform a self--consistent 
treatment of the evolution of the anisotropy of the stellar cusp.
Self--consistent integration of the orbital decay yields 
${dN_{\rm ej}}/{da_*}$, and the number of stars that remain bound to $M_1$ as a function 
of $a_*$ is simply
\begin{equation}\label{cuspdens}
\frac{dN_{\rm bd}}{da_*}=\frac{dN_*}{da_*}-\frac{dN_{\rm ej}}{da_*}. 
\end{equation}
We assume that the stellar cusp remains isotropic, i.e. $\langle e_* \rangle=0.667$. 
The $a_*$ domain is then subdivided in intervals $\Delta a_*$, and for each interval a number of 
bound stars $\propto(dN_{\rm bd}/da_*)\Delta a_*$ is generated with angular momentum 
distribution $\propto L_*^2$. We then compute the probability of finding a star at distance 
between $r$ and $r+dr$ from $M_1$ (this is proportional to the time spent at such distance 
along its orbit), and reconstruct the stellar density profile $\rho(r)$. 
Figure \ref{fig7} shows the profile before and after binary shrinking. The cusp is 
eroded between $\sim 0.01 a_0$ and $\sim 2a_0$, depending on $e_0$.
For $r\lta a_0$, a SIS is flattened to $\rho\propto r^{-0.7}$, while a $r^{-1.5}$ 
cusp becomes $\rho(r)\propto r^{-0.5}$. Such results are independent of the mass ratio $q$. 

As three-body interactions and stellar ejections tend to circularize the orbits of ambient
stars, one may wonder about the validity of the assumption that the cusp remains isotropic. 
We checked that even assuming that all stars still
bound to the binary at the end of the process were set on circular orbits
(i.e., maximum tangential anisotropy), then the cusp slope 
would be flattened at most by a factor $\simeq 0.3$ (for $\gamma=2$, and 
$q=1/729$) with respect 
to the isotropic case. Differences in the cusp slope $\lesssim 0.1$ are common for initially 
shallower cusps and/or larger mass ratios.
We conclude that, depending on $\gamma$ and on the final anisotropy of the bound stars, 
the slingshot mechanism creates a central core as flat as $\rho(r)\propto r^{-(0.4\div 0.8)}$.

\begin{figurehere}
\vspace{0.5cm} 
\centerline{\psfig{figure=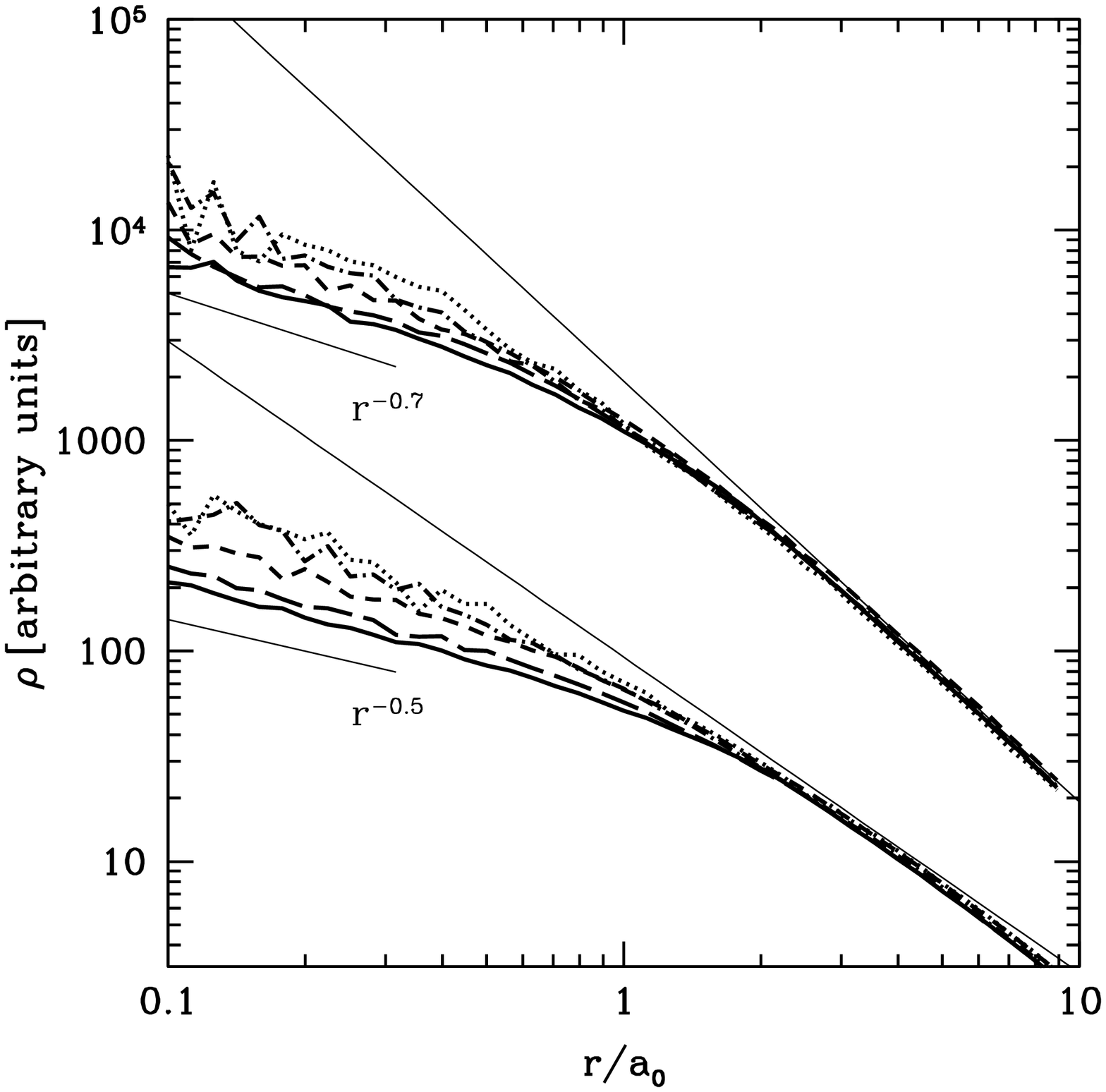,width=3.6in}}
\vspace{-0.0cm}
\caption{\footnotesize The cusp stellar density profiles (in arbitrary units) before
({\it thin lines}) and after ({\it thick lines}) erosion by the shrinking binary. 
{\it Solid line}: $q=1/9$;  {\it Long-dashed line}: $q=1/27$; {\it Short-dashed line}: $q=1/81$; 
{\it Dot-dashed line}: $q=1/243$; {\it Dotted line}: $q=1/729$. 
The upper (lower) set of curves is for $\gamma=2$ ($\gamma=1.5$). 
Fiducial $r^{-0.7}$ and $r^{-0.5}$ power-laws are shown for reference.} 
\label{fig7}
\vspace{+0.5cm}
\end{figurehere}

\subsection{Distribution of kick velocities}

The distribution of stellar ejection velocities in units of the binary circular velocity 
at the initial separation $a_0$, $V_{c,0}$, is shown in Figure \ref{fig8}. The distribution 
cannot be fit by a single or a broken power-law, as in the case of the kick velocities imparted
at fixed binary given separation, as the derivative of the distribution is a monotonic
decreasing function of the ejection speed. While the peak of the distribution shifts toward 
smaller $V/V_{c,0}$ values as $q$ decreases ($V_{\rm peak} \propto \sqrt{q}$), 
the high velocity tail is independent of $q$. For initially eccentric binaries, the 
velocity of $M_2$ at periastron is $>V_{c,0}$ and a significant number of HVSs can then be 
generated. Note that the speed distribution is only weakly dependent on $\gamma$ once $V$ is expressed
in units of $V_{c,0}$.

\subsection{Ejected mass}\label{s5.5.3}

Integration of the kick velocity distribution gives the ejected mass $M_{\rm ej}$. This quantity
is plotted in Figure \ref{fig9} as a function of $q$ for different lower velocity thresholds. 
As discussed above, the total number of ejected stars is approximately $\propto q$ and is
weakly dependent on $e_0$ and $\gamma$. Stellar ejection occurs in a burst lasting from 
few tens to several thousands binary orbital periods (see Fig. \ref{fig10}, note the different 
scale of the time axis in the four panels), i.e. from $10^5$ to $10^7$ yr. The highest 
velocity stars are delayed with respect to the bulk of ejections, as large kicks require 
close binary separations. The ejection rate peaks at earlier times in the case of 
more eccentric binaries, and is larger for steeper cusps. 

\begin{figurehere}
\vspace{0.5cm} 
\centerline{\psfig{figure=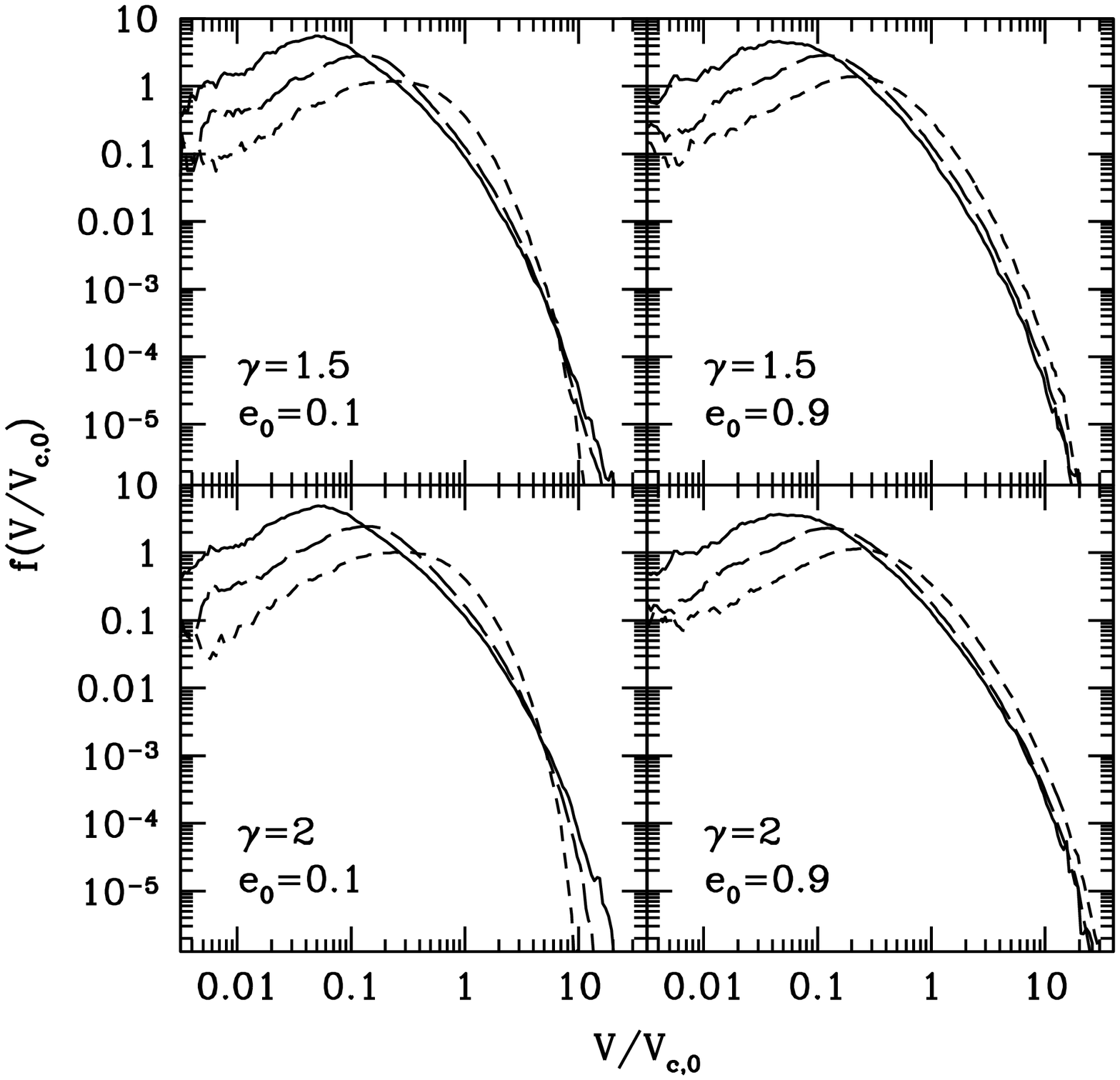,width=3.6in}}
\vspace{-0.0cm}
\caption{\footnotesize Final velocity distribution of ejected stars as a 
function of $V/V_{c,0}$. {\it Short-dashed lines}: $q=1/9$; 
{\it Long-dashed lines}: $q=1/81$; 
{\it Solid lines}: $q=1/729$. The assumed values of $\gamma$ and $e_0$
are listed in each panel.}    
\label{fig8}
\vspace{+0.5cm}
\end{figurehere}

It must be pointed out that the quantity $M_{\rm ej}$ plotted in Figure~\ref{fig9} 
accounts only for the mass ejected by energetic three--body interactions 
occurring in the final stage of the binary evolution. During the entire binary 
evolution, stars in galaxy nuclei are also displaced by the heating associated 
with dynamical friction -- the cumulative effect of many weak encounters with
distant stars. According to Merritt (2006), the total mass displaced by 
dynamical friction depends only weakly on binary mass ratio (for $q$ in the range 
$0.025<q<0.5$), and has little or no effect on the innermost stellar cusp for 
$q<0.1$ (see Fig. 5 in Merritt 2006). 
\begin{figurehere}
\vspace{0.5cm} 
\centerline{\psfig{figure=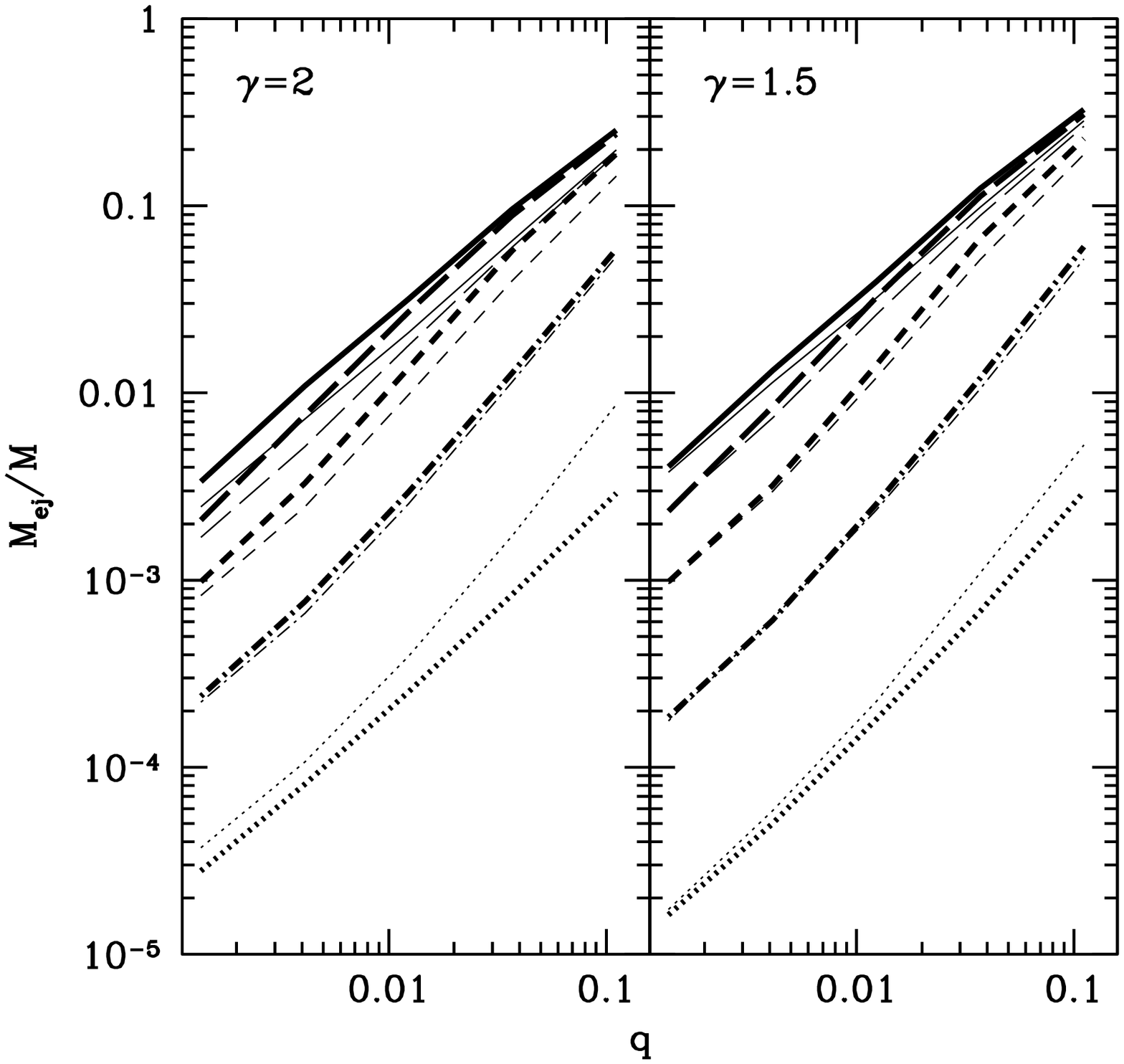,width=3.6in}}
\vspace{-0.0cm}
\caption{\footnotesize Mass ejected from the binary, scaled to the total binary mass, as a function of $q$.
{\it Left panel}: $\gamma=2$. {\it Right panel}: $\gamma=1.5$.{\it Thick lines}: $e_0=0.1$. {\it Thin lines}: 
$e_0=0.9$. In each set of curves lines refers, from top to bottom, to all stars with kick 
velocities $V>0$ (the total mass expelled by the binary) and with $\log{(V/V_{c,0})}>-1+0.5n$ for 
$n=0,1,2,3$.}  
\label{fig9}
\vspace{+0.5cm}
\end{figurehere}
\begin{figurehere}
\vspace{0.5cm} 
\centerline{\psfig{figure=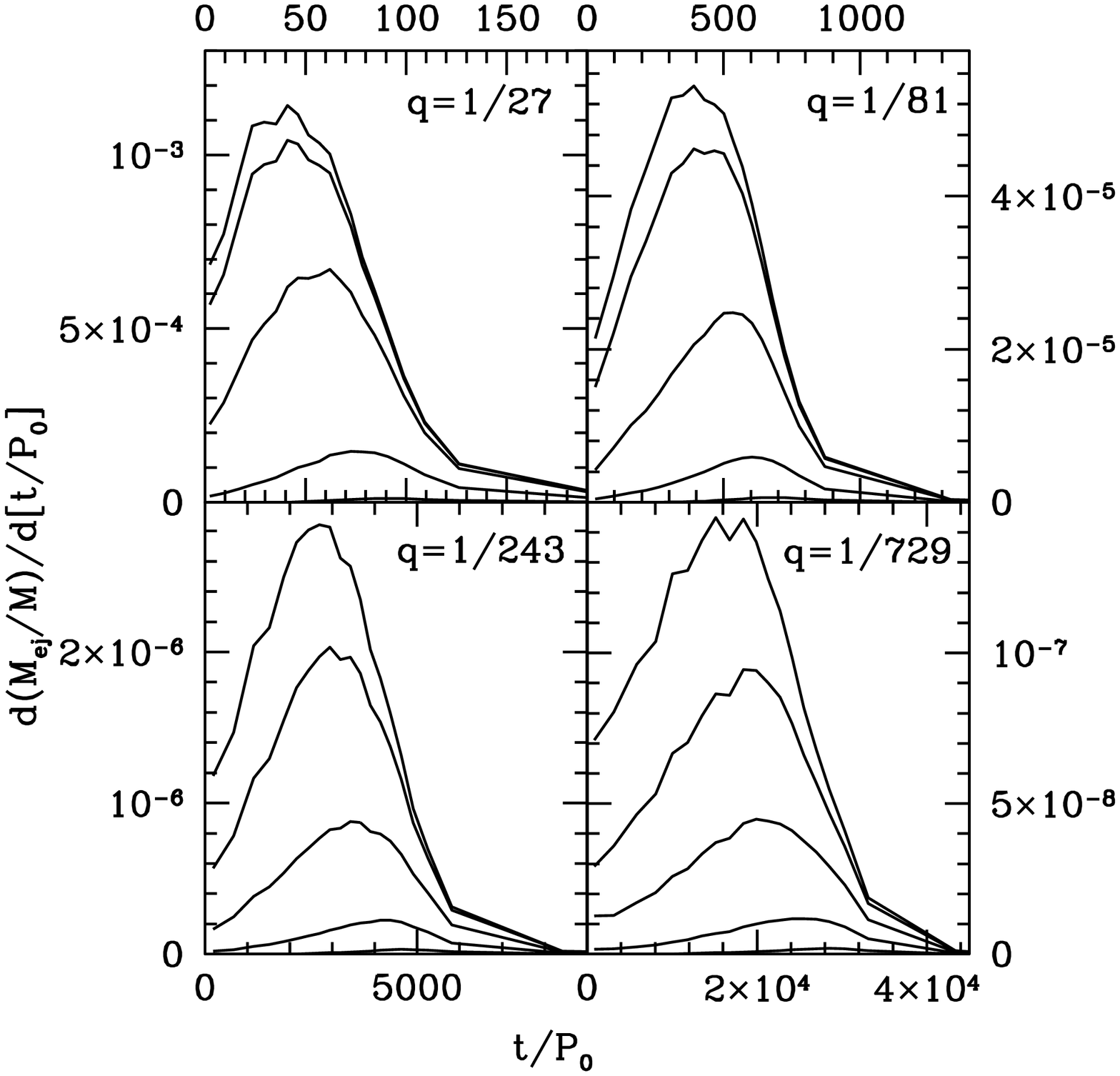,width=3.6in}}
\vspace{-0.0cm}
\caption{\footnotesize Mass ejection rate for binaries with $e_0=0.1$ and different mass ratios embedded
in an SIS. Lines refers, from top to bottom, to all stars with kick 
velocities $V>0$ (the total mass expelled by the binary) and with $\log{(V/V_{c,0})}>-1+0.5n$ for 
$n=0,1,2,3$. Note that the time axis has a different scale in each panel.}  
\label{fig10}
\vspace{+0.5cm}
\end{figurehere}
While the total mass deficit accumulated 
during binary evolution appears to scale with the mass of the binary 
(i.e., to $M_1$ for small $q$), the mass ejected after the energetic encounters discussed 
in this paper, and therefore the number of expected HVSs, scale with the mass $M_2$ of 
the intruder.  The number of stars ejected above a given velocity threshold (in 
units of $V_{c,0}$) is also a weak function of $e_0$ and $\gamma$.

\subsection{Angular properties of HVSs}\label{s5.5.4}

The angular properties of HVSs show several peculiar features, qualitatively similar to 
those discussed in Paper I for the scattering of an unbound stellar population. The ejected
stars are flattened in the binary orbital plane, and, in the case of eccentric binaries, 
are grouped into a ``broad jet'' aligned to the velocity of $M_2$ at periastron.
Both anisotropies are more pronounced in stars undergoing a stronger interaction and
receiving larger kicks. The anisotropy of the HVS population can be quantified by 
computing as a function of binary separation the angles $\langle\theta^2\rangle$ 
and $\langle\phi\rangle$, where $-\pi/2<\theta<\pi/2$ is 
the latitude of the star (the angle between 
the velocity vector $V$ at infinity and the binary orbital plane), and $0<\phi<2\pi$ is its
longitude (the angle between the projection onto the binary orbital plane of the velocity vector at 
infinity and the $x$-axis). For a spherically-symmetric distribution, $\langle\theta^2\rangle
\simeq 0.47$, and $\langle\phi\rangle=\pi$. We find $\langle\theta^2\rangle\sim 0.35-0.40$ for 
all HVSs. As a general trend, 
stars ejected above a given speed tend to become more isotropic 
as the binary shrinks, confirming the analytical result of Levin (2006). 
This effect was already seen in three-body scattering experiments of 
unbound stars (see Paper I for a detailed discussion), 
and is related to the fact that, as the pair decays, its circular 
velocity grows and even stars experiencing relatively weak encounters 
can attain large kick velocities. In terms of longitude, HVSs are ejected almost 
isotropically, though an azimuthal anisotropy become apparent with increasing kick velocities.
The high-velocity tail of the distribution is formed by stars
expelled after a close encounter with $M_2$ near its 
periastron, in a broad jet with $\langle\phi\rangle\simeq3/2\pi$.
As the binary potential is non-Keplerian, such broad jet will precess during binary
evolution on a timescale that depends upon $q$ and $\gamma$ (Levin 2006). 

\section{THE CASE OF SGR A$^*$}\label{sec5.4}

Most of the results presented in the previous sections are scale invariant, i.e. 
they are independent on the absolute value of $M$ and on the chosen normalization of 
the stellar cusp. The only underlying assumption is that the interaction between the 
MBHB and ambient stars begins when the stellar mass inside the binary orbit is equal 
to $2M_2$. 
We are interested in scaling our results to the scattering of stars bound to Sgr A$^*$, 
the massive black hole in the Galactic Center, by an inspiraling companion of intermediate 
mass (Yu \& Tremaine 2002; Sesana, Haardt \& Madau 2007b).  
Let us first express equations (\ref{eq:sis}), (\ref{eq:rbrake}), and (\ref{eq:a0}) in physical units:
\begin{equation}\label{eq:rzero}
r_0=0.43\,\textrm{pc}\,\, (3-\gamma)M_{1,6}\,\sigma_{100}^{-2},
\end{equation}
\begin{equation}\label{eq:rhozero}
\rho_0=1.96\times10^6\msun\,\textrm{pc}^{-3}\,\, (3-\gamma)^{-2}\,M_{1,6}^{-2}
\sigma_{100}^{6},
\end{equation}
\begin{equation}\label{eq:a0phy}
a_0\,=\,0.43\,\textrm{pc}\,\, (3-\gamma)\,M_{1,6}\,\sigma_{100}^{-2}\,q^{1/(3-\gamma)}(1+q)^
{(2-\gamma)/(3-\gamma)},
\end{equation}
where $\sigma_{100}$ is measured in units of $100\,\kms$. 
From $a_0$, the time unit of our experiments is then 
\begin{equation}\label{eq:P_0}
P_0\,=\,1.3\times10^5\,\,\textrm{yr}\,\,(3-\gamma)^{3/2}
M_{1,6}^{1/4}\,q\left(\frac{q}{1+q}\right)^{(2\gamma-3)/(6-2\gamma)}. 
\end{equation}

The stellar density profile around the Galactic Center can be described
as a double power-law, with outer slope $\simeq-2$ and inner slope $\simeq-1.5$ (Genzel 
et al. 2003; Schodel et al. 2007). The massive black hole Sgr A$^*$ weighs $\simeq 
3.5\times10^6\msun$ (Schodel et al. 2002; Ghez et al. 2005). Using $M_{1,6}=3.5$, $\sigma_{100}=1$ 
and $\gamma=1.5$, from equations (\ref{eq:rzero}) and (\ref{eq:rhozero}) we obtain 
$r_0= 2.26\,$pc, and $\rho_0=7\times10^4\,\msun$ pc$^{-3}$, in good agreement with the most recent 
observations (Merritt 2006; Schodel et al. 2007). For the relevant values of $q$ and 
$\gamma$, the typical timescale for orbital decay ranges between $\sim 10^5$ and
$\sim 10^7$ yrs (see eq. \ref{eq:P_0}). In the following, we will consider 
two different mass ratios for the putative MBHB at the Galactic Center, $q=1/243$ and $q=1/729$, 
corresponding to an inspiraling IMBH of mass $M_2\simeq 1.4\times 10^4\msun$ and 
$M_2\simeq 4.8\times 10^3\msun$, respectively, as well as two different slopes for the 
initial inner stellar cusp, $\gamma=1.5$ and $\gamma=1.75$. We will also study the impact 
of binary initial eccentricity. The parameters of the different models considered are listed
in Table 2.

\begin{tablehere}
\begin{center}
\begin{tabular}{|ccccccc|}
\hline
$\gamma$ & $r_0$ & $\rho_0$ & $q$ & $a_0$ & $P_0$ & $V_{c,0}$\\
$$ & $[\rm{pc}]$ & $[\rm{\msun\, pc^{-3}}]$ & $$ & $[\rm{pc}]$ & 
$[\rm{yr}]$ & $[\rm{\kms}]$\\
\hline
   1.5&    2.25&   7.1$\times10^4$&  1/243&  0.058 & 1344& 510\\
   &           &                 &  1/729&  0.028 &  448& 735\\
\hline
   1.75&   1.88&   $10^5$&  1/243&  0.023 & 340&   806\\
   &           &         &  1/729&  0.010 &  91&   1250\\
\hline
\end{tabular}
\end{center}
\caption{\footnotesize Parameters of two different models for the stellar cusp at the 
Galactic Center. The quantities $\gamma, r_0, \rho_0, q, a_0, P_0,$ and $V_{c,0}$ are, respectively,
the cusp slope, the cusp characteristic radius, the density at $r_0$, the binary mass ratio, 
the binary separation at which gravitational slingshots start, the binary orbital period at $a_0$, 
and the binary circular velocity at $a_0$.}
\label{tab:3}
\end{tablehere}

The dynamical evolution of a putative IMBH-Sgr A$^*$ binary is displayed in Figure
\ref{fig20} for $e_0=0.1$. The two upper panels show the time-changing semi-major axis 
$a$ and eccentricity $e$. The former shrinks to $10^{-2}-10^{-3}$ pc in 2-15 Myrs, depending 
on $q$ and $\gamma$. Lighter IMBHs reach smaller separations on a longer timescale.
Our results for the case $\gamma=1.75$ can be directly compared to the 
numerical simulations of Baumgardt et al. (2006) and Matsubayashi et al. (2007), and 
are found to be in excellent agreement. The eccentricity increases rapidly 
in all cases to values $\gtrsim0.8$. We have checked that, when $e_0>0.3$, the binary 
eccentricity can reach values as large as $e\gtrsim 0.95$. This is again in agreement 
with the results of Baumgardt et al. (2006) and Matsubayashi et al. (2007). 
In the figure we have marked with an horizontal dotted line the
separation at which the binary can coalesce in 1 Gyr because of GW emission 
(eq. \ref{eq:agwbound}). Only for a cusp as steep as $\gamma=1.75$ does the pair
actually reach such separation. One should note that, while in our hybrid model 
the eccentricity evolves smoothly, in a realistic situation it will undergo 
discontinuous ``jumps" triggered by rare close encounters, which could induce 
extreme eccentricities and accelerate coalescence. The two lower panels of Figure~\ref{fig20}
show the resulting stellar density profiles after binary erosion. Cusps are flattened 
to $\rho\propto r^{-0.7}$ in the central few$\times 10^{-2}$ pc. The numerical 
simulations of Baumgardt et al. (2006) and Matsubayashi et al. (2007) produce somewhat 
shallower slopes, a discrepancy that may be associated with our assumption 
of an isotropic stellar cusp after the interaction. 

\begin{figurehere}
\vspace{0.5cm} 
\centerline{\psfig{figure=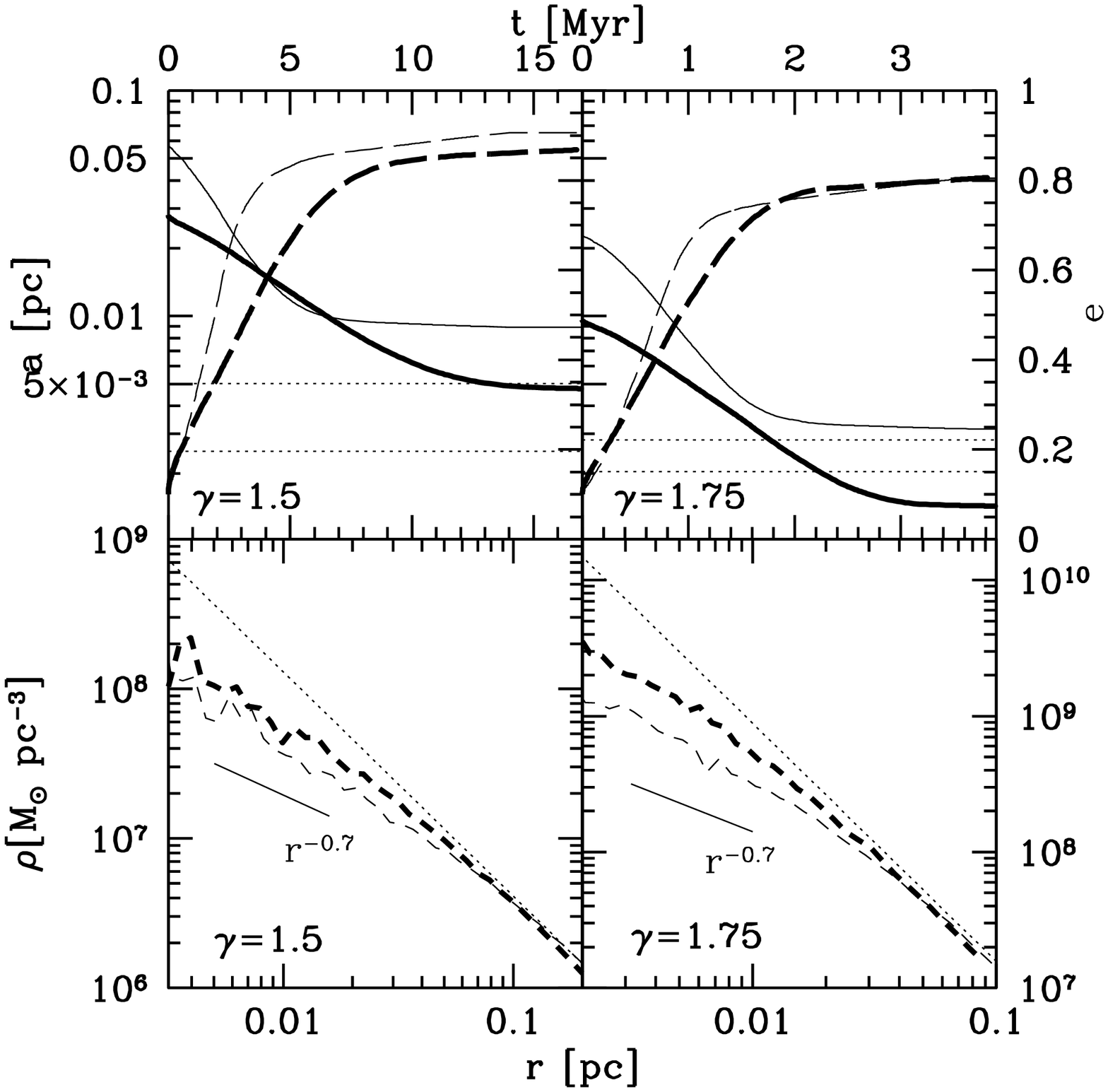,width=3.4in}}
\vspace{-0.0cm}
\caption{\footnotesize The case of Sgr A$^*$. {\it Upper panels}: time evolution of binary semi-major axis 
({\it solid lines}, left axis scale) and eccentricity ({\it dashed-line}, right axis scale)
for $e_0=0.1$. The dotted lines mark the 
separation at which binaries can coalesce because of GW emission in $< 1$ Gyr. 
{\it Lower panels}: evolution of the stellar density profile. In all panels, the 
thin lines are for $q=1/243$, the thick lines for $q=1/729$.
}
\label{fig20}
\vspace{+0.5cm}
\end{figurehere}

The heating of the cusp results in the creation of a population of HVSs. 
This is of particular interest since the discovery of the first HVS in the Milky Way 
(Brown et al. 2005), and an IMBH inspiral onto Sgr A$^*$ is regarded as a possible source 
of hyper--velocity ejections (though the tidal break--up of close binaries by Sgr A$^*$
seems to be supported by observations and statistical studies, see e.g. Perets 2007, Sesana et al. 2007b).  
In Figure~\ref{fig18} we plot the stellar ejection rates as a function of 
time for different models and for different velocities at the radius of influence of Sgr A$^*$.
Modelling the Milky Way potential as the sum of a luminous 
component (Miyamoto \& Nagai 1975) and a dark matter halo (Widrow \& Dubinski 2005), we find 
an escape velocity from the Milky Way of $\simeq 840$ km/s at $r_{\rm inf}$.
This velocity threshold translates, in such a gravitational potential, into about
$450\,\kms$ 10 kpc away from Sgr A$^*$. Stars with $V>300\,\kms$ at $r_{\rm inf}$ 
do not leave the bulge, while stars with $V>600\,\kms$ can reach $4$ kpc away from the 
Galactic Center. Stars with $V>900\,\kms$ and $V>1200\,\kms$ are not bound to the Milky Way
and, at a reference distance of $10$ kpc, have still velocities of $600\,\kms$ and 
$1000\,\kms$, respectively. Figure~\ref{fig18} shows that the ejection occurs in 
a relatively short burst lasting a few 
Myrs, with higher velocity stars being produced at increasingly later times. 
At peak, the ejection rate of HVSs with $V>900\,\kms$ varies between $5\times10^{-5}$ and $2\times
10^{-3}$ yr$^{-1}$, depending on $q$ and $\gamma$. The rate is larger in steeper cusps, 
as stars are more centrally concentrated and are scattered when the binary separation is 
smaller and orbital velocity larger. 

\begin{figurehere}
\vspace{0.5cm} 
\centerline{\psfig{figure=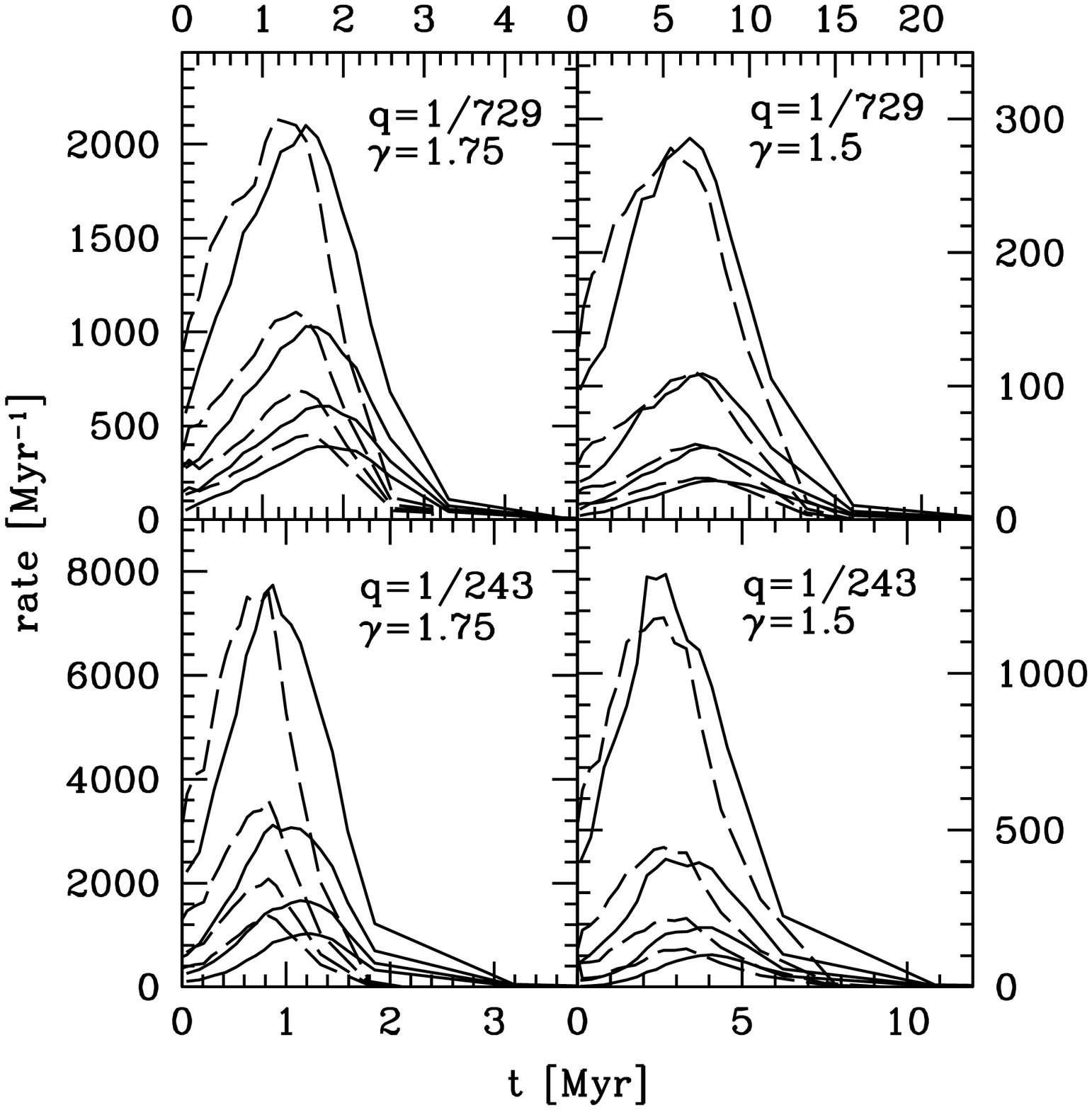,width=3.4in}}
\vspace{-0.0cm}
\caption{\footnotesize Ejection rate of HVSs for an IMBH falling onto Sgr A$^*$. {\it Solid lines}:
$e_0=0.1$; {\it dashed lines}: $e_0=0.9$. For each set 
of curves, lines from top to bottom correspond to a velocity threshold $V>300,\,600,\,900$, 
and $1200\,\kms$ at the radius of influence of Sgr A$^*$, respectively.} 
\label{fig18}
\vspace{+0.5cm}
\end{figurehere}

The speed distribution of HVSs depends on the details of the model. In the range 
$300-1000\,\kms$ (the velocity range of the HVSs observed by Brown et al. 2006), and at 
a Galactocentric distance of 55 kpc (the average distance of the observed HVSs), the distribution 
can be approximated by a power-law $f(V) \propto V^{-1.5}$, almost independent of $q$, $e_0$,
and $\gamma$. Assuming $m_*=1 \msun$, we find a predicted number of HVSs with 
$V>840\,\kms$ at $r_{\rm inf}$ $N_{\rm HVS} \simeq 525\, (1290)$ for $q=1/729$ and 
$\gamma=1.5\, (1.75)$. This number roughly doubles for $q=1/243$ and is
fairly independent on $e_0$. Our results are consistent with Baumgardt et al. (2006), who 
estimate $N_{\rm HVS} \sim 1700$ for $M_2=10^4\msun$, and $N_{\rm HVS} \sim900$ for 
$M_2=3\times10^3\msun$. A peak of ejection occurs after $\sim 1-2$ Myr. Levin (2006) 
finds comparable numbers of stars expelled, and a similarly peaked ejection rate. 

\section{SUMMARY}

We have performed, for the first time, scattering experiments between a MBHB and stars 
drawn from a cusp bound to the primary hole. We have studied the dynamics of the pair
and its orbital decay by three-body interactions, the impact of the gravitational slingshot 
on the stellar density profile, the properties of the ejected stellar population, and 
have scaled our results to the case of Sgr A$^*$. Our results can be quickly summarized 
as follows:

\begin{enumerate}

\item The extraction of the cusp binding energy causes the binary to shrink by a larger 
factor compared to the scattering of unbound stars. The effect is more noticeable in the 
case of small mass ratios $q$.

\item The binary orbital eccentricity increases much more rapidly compared to the unbound 
case. The eccentricity growth is more pronounced in small mass-ratio binaries, and for 
shallower stellar cusps.

\item The combined effects of enhanced orbital decay and eccentricity growth lead very 
unequal-mass binaries to the gravitational wave coalescence phase. The detailed fate of the 
pair depends on the absolute value of its mass. More massive binaries decay faster. 

\item The stellar cusp is eroded, and the total mass removed 
by strong three--body encounters is 2-to-4 times the mass of the secondary hole. While the mass
deficit caused by dynamical friction in a merger event scales with the binary mass 
(and involves mostly distant stars), the mass of stars ejected from the inner cusp by 
highly energetic interactions scales with $M_2$. 
Ejection occurs in a ``burst" lasting from few tenths to several thousands binary orbital periods, 
depending upon $q$.

\item Scaled to the scattering of stars bound to Sgr A$^*$ by an inspiralling IMBH,
our results imply the formation of a core of 0.1 pc in 1-10 Myrs, as well as the 
ejection of 500-2500 HVSs moving with speeds sufficient to escape the gravitational field 
of the Milky Way. In Sesana et al. (2007b) we have used the Brown et al. (2007) sample 
of unbound and bound HVSs together with numerical simulations of the propagation of HVSs in the
Milky Way halo to constrain this ejection mechanisms, and shown that it appears to produce 
a spectrum of ejection velocities that is too flat compared to the observations. 
Future astrometric (as, e.g., {\it GAIA}) and deep wide-field (as, e.g., {\it LSST}) surveys 
should unambiguously identify the ejection mechanism of HVSs, and probe the Milky Way
potential on scales as large as $200$ kpc (Gnedin et al. 2005; Yu \& Madau 2007). 

\end{enumerate}

\acknowledgments
\noindent
Support to this work was provided by NASA grant NNG04GK85G (P.M.). We thank the 
anonimous referee for his/her useful comments that helped to improve the 
quality of the paper.

{}

\end{document}